\documentclass[12pt]{article}
\usepackage[a4paper,margin=24mm]{geometry}
\usepackage{amsmath,amssymb}
\usepackage{dsfont}
\usepackage{amsrefs}
\usepackage{pstricks}
\newcommand{\bin}[2]{\setlength{\arraycolsep}{0pt}
\left(\begin{array}{c}#1\\#2\end{array}\right)}
\newcommand{\clg}[1]{\lceil{#1}\rceil}
\newcommand{\flr}[1]{\lfloor{#1}\rfloor}
\alph{enumi}\roman{enumii}
\def\bn{\begin{enumerate}}\def\en{\end{enumerate}}
\def\bi{\begin{itemize}}\def\ei{\end{itemize}}

\def\C{\mathbb{C}}
\def\Z{\mathbb{Z}}
\def\TL{{\cal T}\!\!_{L}}
\def\LOT{\textsc{lot}}
\def\al{\alpha}
\def\bt{\beta}

\def\sg{\sigma}

\def\be{\begin{equation}}
\def\ee{\end{equation}}
\def\mt{\moveto}

\def\rmt{\rmoveto}
\def\rlt{\rlineto}

\def\rct{\rcurveto}
\def\cput{\rput{0}}
\def\monoid{\rct(0,1)(2,1)(2,0)\rmt(-2,2.5)\rct(0,-1)(2,-1)(2,0)\rmt(2,-2.5)}
\def\wgl{\rct(0,1)(2,1)(2,2)\rmt(0,-2)}
\def\Id{\mathds{1}}
\def\P{{\rm P}}\def\F{{\rm F}}\def\R{{\rm R}}\def\L{{\rm L}}
  \definecolor{lgr}{rgb}{.5,.9,.5}
    \definecolor{ylw}{rgb}{.95,.9,0}

\newcommand{\md}[1]{\put(#1){\mnd}}
\newcommand{\rng}[2]{{[#1,\,#2]}}
\begin{document}
\author{Bernard Nienhuis, 
\\ {\normalsize DITP \& Lorentz Institute for theoretical physics}
\\{\normalsize  P.O. Box 9506, NL-2300 RA Leiden, The Netherlands}
\\{\normalsize \tt nienhuis@ilorentz.org}
\and
Onno E. Huijgen
\\{\normalsize  Donders Centre for Neuroscience, Radboud University}
\\{\normalsize P.O.Box 9101//066, 6500 GL Nijmegen, The Netherlands}
\\{\normalsize \tt o.huijgen@donders.ru.nl} }
\title{The local conserved quantities of the closed XXZ chain }%\\ in terms of the Temperley-Lieb algebra}
\maketitle 
\begin{abstract}
Integrability of the XXZ model induces an extensive number of conserved quantities.  In this paper we give a closed form expression for the series of local conserved charges of the XXZ model on a closed chain with or without a twist.   We prove that each element of the series commutes with the Hamiltonian.
\end{abstract}
\section{Introduction}
% In statistical physics, one is often interested in computing the dynamics of a model.  For exactly solved models,  With that information, one can formulate the Gibbs ensemble and compute the dynamics of a system.  Models where this is possible are called exactly solved models (or quantum integrable models).  While no rigorous definition exists, it is widely accepted that such models are characterized by an extensive number of local conserved quantities~\cites{Caux_2011, quasilocal}.

It is a pleasure to dedicate this article to Rodney Baxter at whose 80-th
birthday party some of this material was presented.  We follow a method which
Rodney has shown us: by direct computation we collect secure knowledge
on the first so many terms in a infinite sequence.  Based on this we
formulate conjectures on the properties of all terms.  When these
conjectures suffice to predict all terms in the sequence, we try to
prove them.  Because we think this recipe can be useful in other
instances, we present not only our results, but also the path that took
us there.

Exactly solved models are a class of models in statistical physics for which the dynamics and/or thermodynamics is computed exactly.   Quantum models that can be solved exactly in this way are called quantum integrable.  While no universally accepted definition of quantum integrability exists, it is widely accepted that it is characterized by an extensive number of local conserved quantities~\cites{Caux_2011, quasilocal}.  By local we mean that the observables involve a finite range of site operators summed over all translations of this range.  While there exist procedures to generate sequences of these quantities, they are computationally expensive.

The one-dimensional Heisenberg model is such a solvable model, as are its spin-anisotropic variants, the XXZ and XYZ models.
We provide a closed form of the sequence of local conserved quantities for the spin-$\frac{1}{2}$ Heisenberg XXZ chain with (quasi-)periodic boundary conditions.  This result is not precisely equal to logarithmic derivatives of the transfer matrix nor to the quantities resulting from the  repeated commutator with a boost operator~\cite{boost}.  But the terms in our sequence can be written as linear combination of the terms of either of these series, up to the corresponding order.

The Hamiltonian of the Heisenberg XXZ chain can be written as a sum of  generators of the Temperley-Lieb (TL) algebra.  There are a number of other models that have a similar relation to the TL algebra, both one-dimensional quantum chains and two-dimensional statistical models.  Well-known examples are the Potts model and the six-vertex model~\cite{6vertex}, as first noted in~\cites{BaxterEquivalence, TLPotts}.  
The $q$-state Potts model~\cite{WuPotts} describes $q$-state particles on a lattice with a fully S$_q$-symmetric interaction.  As a well known generalization of the Ising-model it has many applications in physics, typically on regular lattices.  As a consequence of universality~\cites{universality1,universality2} this model family gives an accurate description of the critical behavior at the phase transition in numerous systems.
But also as a statistical tool it has applications in many fields, e.g. in genetics~\cite{AgrDom}.
The six-vertex is the subject of numerous studies, and among its many applications are predictions concerning the faceting and roughening of crystal surfaces~\cite{roughening} and the shape of crystals in equilibrium with a fluid phase~\cite{CrystalShape}. A central role in this connection is played by the completely packed loop (CPL) model, a lattice gas of loops which pass all edges of the (square) lattice and visit every vertex twice, without intersecting, see e.g. \cites{BaxterEquivalence, Pasquier-ADE1}.  The phrase CPL model was used to distinguish it from the fully packed loop (FPL) model, in which every vertex is visited once.
A very powerful summary of many equivalences between such models is given in~\cites{Pasquier-ADE1, Pasquier-ADE2}.  It gives a wide class of statistical models on a square lattice, in which the vertices take as values the nodes of a graph, such that neighboring vertices take neighboring nodes as values.  The connection with the TL algebra is via the CPL model. 

An important application for conserved quantities in exactly solved models is the formulation of the Generalized Gibbs Ensemble (GGE)\cites{Pozsgay_2013, Fagotti_2013} for the study of the time evolution after a quantum quench.  In standard thermodynamics, isolated systems are almost equivalent to a Gibbs ensemble in which the temperature and chemical potential are chosen to match the internal energy and particle number.  For integrable systems it is expected that the same principle applies, with the Gibbs ensemble replaced by the GGE, in which all or many conserved quantities are conjugated with a thermodynamic potential.
This issue is the more interesting, as a growing number of physical realizations have been achieved for 1D quantum systems, including integrable ones~\cites{GGE, UltracoldMBP, OnedBosons}.
For many quantum integrable models, it has been shown that the GGE does indeed accurately describe the system after a quantum quench~\cites{BosonGGE1, BosonGGE2, QuenchXXZ, GGE}.  It appears, however,  not always sufficient to take the local conserved quantities, as demonstrated in~\cites{XXZQuench,GGEPozsgay}, where unitary time evolution using the Heisenberg XXZ chain Hamiltonian was applied to the N\'eel state.  The existence of additional conserved charges might yield a GGE giving correct predictions~\cites{quasilocal,LocalQuasilocal}. 

A description of the local conserved quantities in the isotropic Heisenberg or XXX chain~\cite{IsotropicHeisenberg} has been given.  The conserved quantities were constructed using a boost operator, and from that a description in terms of polynomial spin functions is created.  The authors noted that``It is an interesting question whether this construction can be generalized to the anisotropic case or to other integrable spin chains''.  Here we extend this result by presenting the conserved quantities of the XXZ chain.

After we completed this investigation, we were alerted to an important parallel publication \cite{Nozawa-Fukai} by the authors, Nozawa and Fukai. They give a construction of local conserved quantities of the closed XYZ chain.    In the conclusion of this paper we discuss the interest of our approach in the circumstance that formally our results can be obtained by a special case of the construction of Nozawa and Fukai.

\section{The Hamiltonian and conserved quantities}
The Hamiltonian of the XXZ model on a periodic chain we write as
\be H = -\frac{1}{2}\sum_{j=1}^L \left(
S_j^{\rm x}\;S_{j+1}^{\rm x}+
S_j^{\rm y}\;S_{j+1}^{\rm y}-\frac{\tau}{2}\;
(S_j^{\rm z}\;S_{j+1}^{\rm z}-1) \right)\label{xxz-H},\ee
where $S^\al_j$ are Pauli matrices acting in the $j$-th factor of ${\cal H}_L=(\C^2)^{\otimes L}$.
For now we choose periodic boundary conditions, $S^\al_{L+1} \equiv S^\al_{1}$.

It is convenient to write  $\tau=-q\!-\!q^{-1}$ and introduce the so-called
monoids, $e_j\equiv e(j,j\!+\!1)$, where
\be e(j,k) \;=\;\frac{1}{2}\left(S_j^{\rm x}\;S_{k}^{\rm x}+
S_j^{\rm y}\;S_{k}^{\rm y}\right)+\frac{q\!+\!q^{-1}}{4}\;(
S_j^{\rm z}\;S_{k}^{\rm z}-1) +
\frac{q\!-\!q^{-1}}{4}(S_j^{\rm z}- S_{k}^{\rm z}) \label{XXZmonoid}\ee
the Hamiltonian, rewritten in terms of the monoids, is then \be H = -\sum_{j=1}^L e_j \label{TL-H}.\ee 
The seemingly unnecessary last term in (\ref{XXZmonoid}) cancels in (\ref{TL-H}) but serves to give the monoids nice properties, as generators in the Temperley-Lieb (TL) algebra, shown in the next section.

As usual in integrable models, the Hamiltonian is the logarithmic derivative of a transfer matrix $\TL(z)$, which has the property $[\TL(z), \TL(z')] = 0$, for different values of the spectral parameter $z$.   The transfer matrix is typically constructed as 
\be \TL(z) = {\rm Tr}_{\rm a} \prod_{j=1}^L R_{{\rm a},j}(z)
\label{TM} \ee
where the so-called R-matrix $R_{j,k}$ acts as the identity in all but the $j$-th and the $k$-th factors, in the space ${\cal H}_{L+1}$, which has one  extra {\it auxiliary} factor $\C^2$.   The standard form (see e.g. \cite{twisted-XXZ}) for $R_{j,k}$ is:
\be R^{\rm st}_{j,k}(z)\;= \;\tfrac{1}{2}\Big([q z]+[z]\Big)\Id + \tfrac{1}{2}\Big([q z]-[z]\Big) S_j^{\rm z}\;S_k^{\rm z} + 
\tfrac{1}{2}[q]\left( S_j^{\rm x}\;S_k^{\rm x}+ S_j^{\rm y}\;S_k^{\rm y}\right)
\label{standardR-M} \ee 
Here and throughout, we use the shorthand $[x]:=x-x^{-1}$.
The Hamiltonian is the logarithmic derivative of the transfer matrix 
\be H\;=\; \left.\frac{z}{\TL(z)}\frac {\partial \TL(z)}{\partial z} \right|_{z=1}\equiv \left.\frac {\partial \log \TL(z)}{\partial \log z} \right|_{z=1}.\label{HfromT} \ee
As a consequence of the mutual commutation of $\TL(z)$ with different values of $z$, the  sequence of higher derivatives 
\be \left.\frac {\partial^n \log \TL(z)}{(\partial \log z)^n} \right|_{z=1} \ee
all commute with $H$ and with each other, i.e. they are conserved quantities.   
Their existence is well-known, but their precise form is not, except for the first few of the series.
% In this paper we present a compact and effective closed form.
\section{The Temperley-Lieb algebra}\label{TL-alg}
The TL algebra is generated by a sequence of generators,  {$e_j$} satisfying the following rules
\be e_j^2 = \tau\, e_j,\qquad e_j\, e_{j\pm 1}\,e_j = e_j,\qquad
  [e_j,e_k]=0\mbox{ for }|j-k|>1
\label{TL-rules}\ee
For the standard TL algebra, with an open sequence of generators, this description is complete.
In the affine version, applicable to the periodic XXZ chain, the indices are read modulo $L$.
For the full description of the periodic chain, we take the extended affine
TL algebra\cite{XXZ-TL} with an additional generator $\rho$ and its inverse $\rho^{-1}$, satisfying:
\be \rho\, e_j = e_{j+1}\,\rho, \qquad\text{ and }\qquad (e_1\,\rho)^{L-1} = e_1\,\rho^{L+1} \label{TL-affine}\ee
Of course, with the introduction of $\rho$ and $\rho^{-1}$ and the first rule of (\ref{TL-affine}) the monoids are no longer independent, and it suffices to keep as generators only $\{e_1, \rho, \rho^{-1}\}$.

%
%   GRAPHICAL TEMPERLEY-LIEB RULES
%
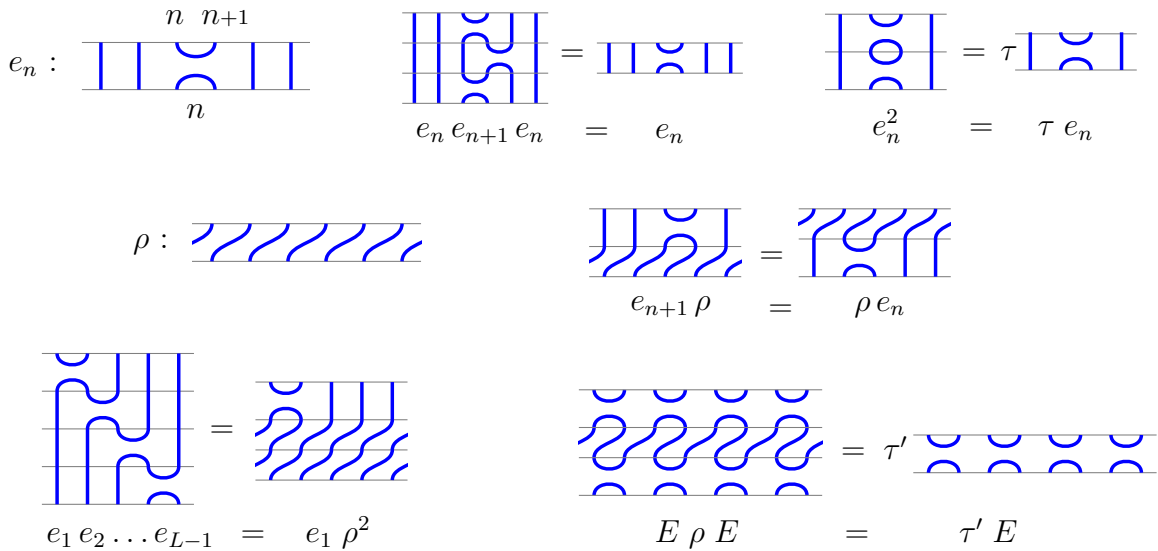
\begin{figure}[b]
  \setlength{\unitlength}{1cm}\begin{picture}(15,8)(0,.5)
    \def\hln{\pscustom[linecolor=gray,linewidth=.4pt]}
    \def\crvs{\pscustom[linecolor=blue,linewidth=1.3pt]}
	\def\nxtl{\rlt(0,2.5)\rmt(2,-2.5)}
    \put(.5,6.5) {\psset{unit=2.5mm}
	\raisebox{2.5mm}{$e_n$ : }
	\pspicture(1,0)(13,2)
	\put(5.4,3.5){$n$}
	\put(7.3,3.5){$n${\scriptsize +1}}
        \put(6.5,-1.5){$n$}
%        \put(-1,3.5){\scriptsize spin labels:}
%        \put(-1,-1.5){\scriptsize monoid labels:}
	\hln{
	\mt(1,2.5)\rlt(12,0)\mt(1,0)\rlt(12,0)}
	\crvs{
	\mt(2,0) \nxtl\nxtl
	\mt(6,0) \rct(0,1)(2,1)(2,0)
	\mt(6,2.5)\rct(0,-1)(2,-1)(2,0)
	\mt(10,0) \nxtl\nxtl}\endpspicture}
   \put(5.7,6) {
	\psset{unit=1.6mm}
	\pspicture(0,-2)(30,8)
	\crvs{
	\mt(0,0)\nxtl\nxtl\monoid\nxtl \nxtl%\nxtl 
	\mt(0,2.5) \nxtl\nxtl\nxtl\monoid \nxtl%\nxtl
	\mt(0,5) \nxtl\nxtl\monoid\nxtl \nxtl}
	\hln{
          \mt(15,2.5)\rlt(12,0)\mt(15,5)\rlt(12,0) 
	\mt(-1,2.5)\rlt(12,0)\mt(-1,5)\rlt(12,0) 
	\mt(-1,7.5)\rlt(12,0)\mt(-1,0)\rlt(12,0)}
	\crvs{\mt(16,2.5) \nxtl\nxtl\monoid \nxtl\nxtl}
	\cput(13,4){=}
	\cput(5,-2.5){ $e_n\,e_{n+1}\,e_n$}\put(14,-3){=}
	\cput(21,-2.5){$e_n$}\endpspicture}
\put(11.3,6){
        \psset{unit=2mm}
        \pspicture(0,-2.5)(18,5)
        \hln{\mt(-1,0)\rlt(8,0) \mt(-1,2.5)\rlt(8,0) \mt(-1,5)\rlt(8,0)
          \mt(11.5,1.3)\rlt(8,0) \mt(11.5,3.7)\rlt(8,0)}
	\crvs{
	\mt(0,0)\nxtl\monoid\nxtl\rmt(-8,2.5)\nxtl\monoid\nxtl
	\mt(12.5,1.3)\nxtl\monoid\nxtl}\put(8,2){$=\,\tau$}
	\put(2.1,-3){$e_n^2$}\put(8.5,-3){=}\put(13,-3){$\tau\;e_n$}
	\endpspicture}
    \put(8,3.4){
	\psset{unit=2mm}
	\pspicture*(0,-3.1)(10,5.1)
	\cput(5,-2){ $e_{n+1}\,\rho $}
        \hln{\mt(0,2)\rlt(10,0) \mt(0,4.5)\rlt(10,0) \mt(0,0)\rlt(10,0)}
	\crvs{\mt(-1,0)\wgl\wgl\rmt(4,0)\wgl\wgl
	\mt(1,2)\nxtl\nxtl\rmt(4,0)\nxtl}
	\crvs{\mt(3,0)\wgl\wgl\mt(5,2)\monoid}
	\endpspicture 
	\hspace{1mm}\raisebox{8mm}{ =}\hspace{-2mm}\raisebox{1mm}{=} \hspace{-2mm}	
	\pspicture*(0,-3.1)(10,5.1)
	\cput(5,-2){ $\rho \,e_n $}
        \hln{\mt(0,2.5)\rlt(10,0) \mt(0,4.5)\rlt(10,0) \mt(0,0)\rlt(10,0)}
	\crvs{\mt(-1,2.5)\wgl\wgl\rmt(4,0)\wgl\wgl
	\mt(1,0)\nxtl\rmt(4,0)\nxtl\nxtl}
	\crvs{\mt(3,2.5)\wgl\wgl\mt(3,0)\monoid}
	\endpspicture}
    \put(2,4.2) {\psset{unit=2.5mm}
        \raisebox{2mm}{{ $\rho$} : }
        \pspicture*(1,-.1)(13,3.1)
        \hln{\mt(-1,0)\rlt(14,0) \mt(-1,2)\rlt(14,0)}
        \crvs{
        \mt(0,0) \wgl\wgl\wgl\wgl\wgl\wgl\wgl}
        \endpspicture}
\put(1,1){
	\psset{unit=2mm}
	\pspicture(0,0)(8,10)
        \hln{\mt(-1,0)\rlt(10,0) \mt(-1,2.5)\rlt(10,0) \mt(-1,5)\rlt(10,0) 
          \mt(-1,7.5)\rlt(10,0) \mt(-1,10)\rlt(10,0)}
	\crvs{
	\mt(0,0)\nxtl\nxtl\nxtl\monoid
	\mt(0,2.5)\nxtl\nxtl\monoid\nxtl
	\mt(0,5)\nxtl\monoid\nxtl\nxtl
	\mt(0,7.5) \monoid\nxtl\nxtl\nxtl}
	\endpspicture
	\raisebox{9mm}{\quad= } 
	\pspicture*(1,-.1)(11,10.1)
        \hln{\mt(1,1.5)\rlt(10,0) \mt(1,3.5)\rlt(10,0) \mt(1,5.5)\rlt(10,0) \mt(1,8)\rlt(10,0) }
	\crvs{
	\mt(0,1.5) \wgl\wgl\wgl\wgl\wgl\wgl
	\mt(0,3.5) \wgl\wgl\wgl\wgl\wgl\wgl
	\mt(2,5.5) \monoid\nxtl\nxtl\nxtl}
	\endpspicture}
         \put(1, .5){$e_1\,e_2\ldots e_{L-1} \;\;=\quad e_1\;\rho^2$}
    \put(8,1.1) {\psset{unit=2mm}\pspicture*(0,-.1)(16,7.1)\crvs{
        \mt(1,0)\monoid\rm(1,0)\monoid\rm(1,0)\monoid\rm(1,0)\monoid
        \mt(-1,2.5) \wgl \wgl \wgl \wgl \wgl \wgl \wgl \wgl \wgl
        \mt(1,4.5)\monoid\rm(1,0)\monoid\rm(1,0)\monoid\rm(1,0)\monoid}
        \hln{\mt(0,0)\rlt(16,0) \mt(0,2.5)\rlt(16,0)\mt(0,4.5) \rlt(16,0)\mt(0,7)\rlt(16,0)}
\endpspicture}
      \put(12.4,1.4) {\psset{unit=2mm}\pspicture*(0,-.1)(16,7.1)
      \hln{\mt(0,0)\rlt(16,0) \mt(0,2.5) \rlt(16,0) }
     \crvs{\mt(1,0)\monoid\rm(1,0)\monoid\rm(1,0)\monoid\rm(1,0)\monoid}\endpspicture}
    \put(11.45,1.6){$=\;\tau'$}
    \put(9,.5){$E\;\rho\;E$}\put(11.5,.5){=}\put(13,.5){$\tau'\;E$}
    \end{picture}
    \caption[TL-rules]{The TL generators and rules in graphical notation. The top row shows the generators $e_n$ and the rules of the TL algebra.  The middle row shows the additional generator $\rho$ and its defining relation, for the affine TL algebra.  The bottom row shows the rules (\ref{TL-affine}) and  (\ref{TL-ext}).
      In the top-left: note the monoid index  under, and the spin index above the figure.
      \label{TLgraph}}    
\end{figure}

A graphical notation neatly encodes the generators and relations as shown in figure~\ref{TLgraph}.  The operators are represented by  a link pattern of two (cyclic) rows of $L$ points.  Multiplying operators is represented by stacking their diagrams, last one on top.  The product is then defined by how points of the top row and bottom row are linked.
The last rule of (\ref{TL-affine}) is rewritten as $e_1 \,e_2\ldots e_{L-1} = e_1\,\rho^2$ before it is depicted in the figure.  We will make use of one additional rule applicable only for even $L$,
\be E\,\rho\,E = \tau'\,E\qquad\text{ where }\qquad
E := \prod_{j=1}^{L/2} e_{2j-1}\label{TL-ext}\ee
This rule\cite{XXZ-TL} and the first of (\ref{TL-rules}) imply that each closed loop, can be removed under multiplication of the expression by $\tau$ for contractible and  $\tau'$ for non-contractible loops.
In the representation (\ref{XXZmonoid}), $\tau = -q\!-\!q^{-1}$, and $\tau' = 2$.
The results in this paper depend only on the rules (\ref{TL-rules}), and are unaffected by (\ref{TL-affine}) and (\ref{TL-ext}) or the value of $\tau'$.

\section{Definitions and main result}
\begin{figure}[tb]
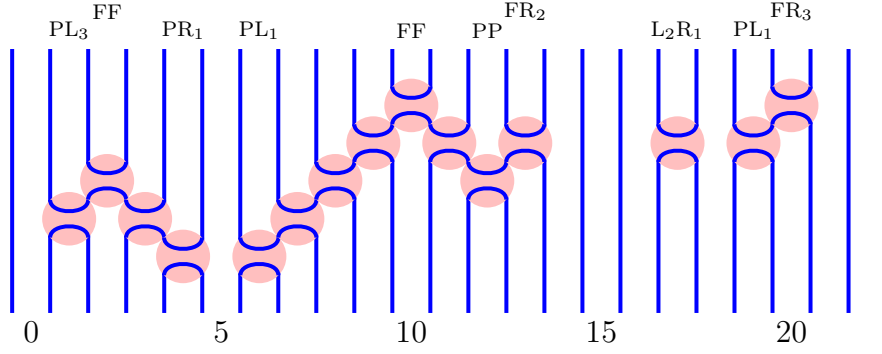

\centerline{\psset{unit=5mm}
\pspicture(0,-.5)(24,8)
\def\mnd{\pscircle*[linecolor=pink](0.5,0.5){.71}
\pscustom[linecolor=blue,linewidth=1.5pt]{
\mt(0,0) \rct(0,.4)(1,.4)(1,0)
\mt(0,1) \rct(0,-.4)(1,-.4)(1,0)}}
\def\vln{\rlt(0,7)\rmt(1,-7)}
\def\ql{\vln\vln\vln\vln}
\pscustom [linecolor=blue,linewidth=1.5pt]{
\mt(0,0) \ql\ql \ql\ql\ql\ql}
\md{1,2}\md{2,3}\md{3,2}\md{4,1}
\md{6,1}\md{7,2}\md{8,3}\md{9,4}\md{10,5}\md{11,4}\md{12,3}\md{13,4}
\md{17,4}\md{19,4}\md{20,5}
\cput(.5,-.5){0}
\cput(5.5,-.5){5}
\cput(10.5,-.5){10}
\cput(15.5,-.5){15}
\cput(20.5,-.5){20}
\cput(1.5,7.5){\scriptsize$\rm PL_3$}
\cput(2.5,8){\scriptsize$\rm FF$}
\cput(4.5,7.5){\scriptsize$\rm PR_1$}
\cput(6.5,7.5){\scriptsize$\rm PL_1$}
\cput(10.5,7.5){\scriptsize$\rm FF$}
\cput(12.5,7.5){\scriptsize$\rm PP$}
\cput(13.5,8){\scriptsize$\rm FR_2$}
\cput(17.5,7.5){\scriptsize$\rm L_2R_1$}
\cput(19.5,7.5){\scriptsize$\rm PL_1$}
\cput(20.5,8){\scriptsize$\rm FR_3$}
\endpspicture}
\caption[Word in TL$_1$]{The word
$p = e_2\,e_1\,e_3\,e_4\,e_{10}\,e_{9}\,e_{8}\,e_{7}\,e_{6}\,e_{11}\,e_{13}\,e_{12}\,e_{17}\,e_{20}\,e_{19}$ in graphical representation.   The index of the monoid runs horizontally, and the order in the product vertically, bottom first.  Only for nearest neighbors the order matters, as more distant monoids commute.   Positions 5, 14, 15, 16 and 18 are vacancies, forming three gaps.   Neighbor pairs of which the right-hand element is placed higher, count as a transposition; like (1,2), (12,13), (19,20) and the four neighbor pairs in the range 6...10.  These sequences we call ascending. A descending sequence, e.g. 10...12 are transposition free. The parameters $(w,t,v,g)$ of $p$ are $(20,7,5,3)$.  For the code in the top of some columns, see section \ref{SecProof}.}\label{wordinTL1}
\end{figure}

In this section we define in the affine TL algebra  a sequence of operators, $Q_k$, with $k\in\mathbb{N}$, of which $Q_1 := -H = \sum_j e_j$.
In order to give a clear description of the $Q_k$, we first define a few terms.
\begin{itemize}
\item A {\em word} is a product of monoids, where the monoids play the role of {\em letters}.
\item Positional references as {\em left} and {\em right} are used for the space of indices of the monoids:  monoid $e_i$ is {\em left} of $e_{i+1}$.
\item The place of a monoid in a word is referred to in temporal terms like {\em before} and {\em after}, {\em preceding} and {\em following}: in $e_i\,e_j$ the monoid $e_i$ comes {\em after} (or {\em follows}) $e_j$.  
\item We call a monoid {\em initial} in a word if it commutes with all the monoids {\em before} it, and {\em final} if it commutes with all the monoids {\em after} it. In other words an {\em initial} monoid can be moved to the first position, and a {\em final} to the last.
\item The {\em length}, $\ell$, of a word is the number of  monoids in the word.
\item {\em Reducing} a word is making it shorter by application of the rules (\ref{TL-rules}).
\item A word which cannot be further reduced is called {\em irreducible}, or, when we wish to emphasize the fact that it resulted from a reduction, {\em reduced}.
\item The {\em width}, $w$, of a word is the difference between the largest and the smallest index of the monoids in the word, incremented by one.
% \item The number of {\em transpositions} $t$ of a word denotes the number of times that an $e_j$ follows (is placed to the left of) $e_{j-1}$. 
\item We define a {\em transposition} as the event that an $e_j$ follows (as operator is applied after) $e_{j-1}$. We denote the number of transpositions in a word by the symbol $t$.
\item The complete reversal of the order of monoids in a word, we will denote as {\em time reversal}.  Under  {\em time reversal} the number of transposition $t$ changes into $w-t-1$.
\item We reserve the word {\em reflection} for spatial reflection, i.e. changing the index of the monoids $i \to m-i$ for some $m$.
\item In a word that contains the monoids $e_i$ and $e_k$ we call the absence of $e_j$, with $i<j<k$, a {\em vacancy}. The number of vacancies in a word is denoted by $v$.
\item A continuous sequence of vacancies we call a {\em gap} and the number of gaps in a word is denoted by $g$.
\item A {\em connected} word is a word without vacancies.
\item Irreducible words with the additional property that no monoid appears more than once will play a special role in this paper.  We therefore introduce the symbol  ${\rm TL}_k$ for the set of words in which no monoid appears more than $k$ times.  
\end{itemize}
Note that TL$_k$  does not span an algebra, as the product of words  $\in {\rm TL}_k$ may involve the same monoid up to $2k$ times.
To illustrate the properties of TL$_1$,  figure~\ref{wordinTL1} shows the graphical representation of a word $\in{\rm TL_1}$.  

Now we are ready to present the central result of this paper.  
We write the symbols $Q_k$ as a linear combination of unique {\em irreducible} words $q$
\be Q_k = \sum_{q} D_k(q)\;q \ee
For all words $q$ in which any monoid occurs more than once, the coefficient $D_k(q)=0$; one may as well limit the sum to TL$_1$:
\be Q_k = \sum_{q \in {\rm TL_1}} D_k(q)\;q, \label{QkinDkq}\ee
Of the remaining words, the coefficient depends on the index $k$, as expected, and is further determined completely by the following properties of $q$: its width, $w_q$, the number of transpositions, $t_q$, the number of vacancies, $v_q$, and the number of gaps, $g_q$.   \be D_k(q) = C_{k}(w_q,t_q,v_q,g_q).  \label{DkinCkwtvg}\ee
Note in particular, that this implies that $Q_k$ is invariant for translations, i.e.  it commutes with the generator $\rho$.

Introducing a shorthand for the coefficient of connected words $C_k(w,t) := C_k(w,t,0,0)$, the dependence on $v$ and $g$ can be written as:
\be C_k(w,t,v,g) = (-\tau)^g \;C_k(w\!+\!v\!+\!g,\;t\!+\!v\!+\!g) \label{CkwtvginCkwt}\ee
and $C_k(w,t)$ is a polynomial in $\tau$:
\be C_k(w,t) = \sum_{j=0}^{(k\!-\!w)/2}\tau^{k-w-2j}\; Z_k(w\!+\!2j,\;t\!+\!j) \label{CkinZk}\ee
where
\begin{equation} Z_{k}(w,t) = \left\{\begin{array}{l}0 \quad\mbox{ if }\quad w>k%,\; t<0 \;\mbox{ or }\;t\geq w
\\ [2mm]
(-1)^{t}\left[
\bin{\clg{\frac{k}{2}}-t-1}{k-w}  \;+\; \bin{\flr{\frac{k}{2}}-t-1}{k-w} \right]\quad\mbox{otherwise}\end{array}\right.  
 \label{eq:Z-table} \end{equation}
With these definitions we can write $Q_k$ as
\be Q_k = \sum_{q\in{\rm TL}_1} C_k(w_q,t_q,v_q,g_q) \,q\,.\label{QkinCk}\ee  
Having thus defined the operators $Q_k$, our main result is the {\it Theorem:}
\be [Q_k,H]\;=\;0 \label{main} \ee
A few words of explanation and attention are in order.

While the coefficients of the words are now well-defined, one may wonder about the complexity of the sum in (\ref{QkinCk}).   Due to the restriction in (\ref{eq:Z-table}) words with $w>k$ are not included as they have zero coefficient.   An allowed word can be completely specified by stating (i) which monoids in a range of $w$ are included, and (ii) which nearest neighbor pairs of included monoids are transposed (i.e. placed in the order $e_{j+1} \,e_j$ rather than  $e_{j}\, e_{j+1}$).   These data completely determine the word, as monoids further apart than nearest neighbor, commute.   The total number of words of length $k$ contributing  to $Q_k$ is exactly $L\;2^{k-1}$:  a factor two for each pair of neighboring monoids (transposed or not) and a factor $L$ for the possible translations.  The total number of words (with non-zero coefficient) we find less than  $L\;2^{k}$.

The parameters $w$, $t$, $v$ and $g$ are all non-negative and not completely independent.   For instance, by definition $t < w\!-\!v\!-\!g$, $g\leq v$ and $v\!+\!g < w$.   These restrictions are not encoded in the above expression for $C_k(w,t,v,g)$, i.e.  the expression for $C_k(w,t,v,g)$ can be non-zero for combinations of the variables that do not correspond with an existing word; they simply do not appear in the sum of (\ref{QkinCk}).   Only for legitimate combinations of its arguments is $C_k(w,t,v,g)$  meaningful.

While the sequence of  $Q_k$ is now defined explicitly, 
there is still considerable freedom to vary on the definitions, while retaining the defining property (\ref{main}).
In particular if all members of the series $Q_k$ commute with $H$ so do their linear combinations.

The $Q_k$ as defined in (\ref{QkinCk})  for odd (even) $k$  are symmetric (anti-symmetric) both under time reversal and under reflection.   But even with that restriction one can still take linear combinations of $Q_k$ with $k$ all even or all odd.   What turns out to be essential in the proof of (\ref{main}) is the {\it triangle equation}:
\be\left\{\begin{array}{lcc}
 Z_k(w,t\!-\!1) + Z_k(w,t) + Z_k(w\!+\!1,t) = 0 &\quad\text{for}\quad& 0 < t < w  
\\[2mm]Z_k(k\!+\!1,t)  = 0
\end{array}\right.\label{triangle}\ee
This equation is satisfied by  (\ref{eq:Z-table}), but in fact, together with (\ref{QkinDkq}-\ref{CkinZk})  it suffices for all $Q_k$ to commute with $H$.   Equation   (\ref{eq:Z-table}) is only a particular closed form example of the solutions of  (\ref{triangle}).
\section{Origin of the result}
In this section, we explain how we arrived at (\ref{CkinZk}-\ref{eq:Z-table}) and  (\ref{triangle}).  While we think this history is illuminating, it is not essential for the understanding of the rest of the paper.

We computed the first ten conserved quantities by calculating the corresponding logarithmic derivatives of the transfer matrix.
This is not the most efficient way to compute conserved operators, but it happened to be instrumental in the task of discovering their regularities.
Studying the resulting expressions we identified a number of their properties, which seemed to be sufficient to define the entire series.

\def\Rtl{R^\textsc{tl}}
\def\cRtl{\check{R}^\textsc{tl}}
In order to make use of the simplifying properties of the TL algebra, we took a definition of the R-matrix alternative to (\ref{standardR-M}): \be \Rtl_{j,k}(z)  \;:= \;X_{j,k}\;\cRtl_{j,k}(z) \;:= \;X_{j,k}\;\Big([q z]\,\Id + [z]\,e(j,k)\Big)\,, \label{TL-Rmatrix}\ee 
where the exchange operator $X_{j,k}$ interchanges the vectors in the $j$-th and $k$-th factor: for any operator $M_{j,k}$ acting non-trivially ony in the $j$-th and $k$-th factor, $M_{j,k}\,X_{j,m}=X_{j,m}\,M_{m,k}$. The transfer matrix $\TL(z)$ is unchanged as $R^{\rm st}_{j,k}(z)$ is replaced by $\Rtl_{j,k}(z)$ in (\ref{TM}):
\be \TL(z) = {\rm Tr_a} \prod_{j=1}^L \Rtl_{{\rm a},j}(z) \ee
We can collect the exchange operators  $X_{{\rm a},j}$ by
\be\begin{split} \TL(z) &=  
{\rm Tr_a} \prod_{j=1}^L X_{{\rm a},j}\;\cRtl_{{\rm a},j}(z)
 \\ &=\; {\rm Tr_a}\; X_{{\rm a},L} \left(\prod_{j=2}^L \cRtl_{{\rm a},j}(z)\; X_{{\rm a},j-1} \right) \cRtl_{{\rm a},1}(z) 
\\ & =\; {\rm Tr_a}\; X_{{\rm a},L} \left(\prod_{j=2}^L  X_{{\rm a},j-1} \; \cRtl_{j-1,j}(z) \right) \cRtl_{{\rm a},1}(z) 
\\ & =\; {\rm Tr_a}\; \left(\prod_{j=1}^L X_{{\rm a},j}\right) \left(\prod_{j=2}^L  \cRtl_{j-1,j}(z) \right) \cRtl_{{\rm a},1}(z) 
\\ & =\; {\rm Tr_a}\; \left(\prod_{j=1}^L X_{{\rm a},j}\right) 
\left(\prod_{j=1}^{L-1}  \Big([q z]\,\Id + [z]\,e_j\Big)\right) \Big([q z]\,\Id + [z]\,e({\rm a},1)\Big)
\\ & =\; {\rm Tr_a}\; \left(\prod_{j=1}^L X_{{\rm a},j}\right) 
\left(\prod_{j=0}^{L-1}  \Big([q z]\,\Id + [z]\,e_j\Big)\right)
\end{split}\ee
where in the last step we simply viewed the auxiliary space as the zero-th factor of  ${\cal H}_{L+1}$.  In all of the steps, the ${\rm Tr_a}$ applies to the complete expression.  The effect of $\prod_{j=1}^L X_{{\rm a},j}$  (a cyclic permutation of the states over the factors  in the tensor product space) is independent of $z$, and therefore disappears in the logarithmic derivative.
We further simplify the calculation of the conserved quantities by first dividing $\TL(z)$ by $[q z]^L$,  and do the differentiation w.r.t.  $b:=[z]/[q z]$ rather than w.r.t.  $\log z$.   This only changes each derivative by a trivial factor and the addition of a linear combination of the lower derivatives.   
The obvious advantage is that each factor $(\Id + b\,e_j)$ admits only one differentiation.
\begin{equation}
  \label{eq:Q_from_T}
  \tilde{A}_k := \left.\left(\frac{\partial }{\partial b} \right)^{k} \log \prod_{j=0}^{L-1}(\Id + b\,e_j) \right|_{b = 0}
\end{equation}
The object $\tilde{A}_k$ is a linear combination of words in the TL algebra, with length $\ell\leq k$.

There is an alternative procedure to construct conserved quantities using a boost operator~\cite{boost}.  The resulting series of conserved quantities differs from $\tilde{A}_k$ (\ref{eq:Q_from_T}) and from $Q_k$ (\ref{QkinCk}).   But the $k$-th member of one series can be written as a linear combination of the first $k$ of one of the others.

From  $\tilde{A}_k$ we constructed  $A_k$ as  
\be A_k = \frac{\tilde{A}_k}{(k\!-\!1)!} + \sum_{i=1}^{k/2} a_{k,i}\;\tilde{A}_{k+1-2i}\,, \label{Atilde2A}\ee 
in which $a_{k,i}$ were chosen such that the resulting $A_k$ was (anti-)symmetric for (even) odd $k$.  These $A_k$ differ from the $Q_k$ defined above, but they were the starting point for a number of observations which eventually led to a conjecture for all $k$.

We observed that the coefficients of connected words only depend on their length and the number of transpositions.   Note that by construction, all words in $\tilde{A}_k$ resulted from reducing a word of length  $\ell = k$ monoids.  Reduction decreases the length of a word in steps either by two via $e_{i}\,e_{i \pm 1}\,e_{i} = e_{i}$, or by one while adding a factor of $\tau$ via $e_i^2 = \tau\, e_{i}$.  The coefficient of words of length $\ell = k-i$ must therefore be an even (odd) polynomial in $\tau$ for even (odd) $i$.   While this argument is valid only for the words in $\tilde{A}_k$, the property is also observed for words in $A_k$.   

For connected words of width $w=k$, we noted that their coefficients were simply equal to $(\!-\!1)^{t}$.  
For shorter connected words the prefactors are polynomials in $\tau$ of degree $k-w$, and we focus on the leading order term  (\LOT), i.e.  the coefficient of $\tau^{k-w}$.   The~\LOT~of a connected word in $A_k$ with $t$ transpositions and width $w$ we call $Z_k(w,t)$.
We observed that $Z_k(k\!-\!1,t) = (t-k/2+1) (\!-\!1)^t $.
For $w = k-2$, the~\LOT s are, up to an overall factor $(\!-\!1)^t$, equal to a quadratic function of $t$, which we recognized as $Z_k(k\!-\!2,t) = (\!-\!1)^t(t+1)(k\!-\!t-2)/2$.   Continuing this pattern, we noted that if two~\LOT s of connected words with $t$ and $t-1$ transposition, respectively, and width $w$, are added, they sum up to minus the~\LOT~for a connected word of width $w\!+\!1$ and $t$ transpositions, i.e. precisely  the first line of equation (\ref{triangle}). This property turns out to hold for all connected words, and thus provides an opening to compute $Z_k(w,t)$ from  $Z_k(w\!+\!1,t)$:  If for a given $k$ all~\LOT s for connected words of some width $w\!+\!1$ are known, we need to infer just one value for a word of width $w$, and then all other~\LOT s for that width can be computed by repeatedly solving the triangle equation  (\ref{triangle}).  As an example table \ref{tab:Z-table} shows the~\LOT~of the connected words for $k=6$ and 7.
\begin{table}[ht]\caption[Z-table]{The tables $Z_6(w,t)$ and $Z_7(w,t)$, i.e. the coefficient of the~\LOT~for connected words.  The values are not those of equation (\ref{eq:Z-table}), but the ones applicable to $A_k$.  Left-right reflection corresponds to the transformation $t\to w-t-1$.  Any up-triangle of three entries, like the shaded triplets of cells, add up to zero. Similarly the sum of pairs of entries at the boundary on every other row vanishes, as the shaded pair in the second table. \label{tab:Z-table} }
  \centerline{
\def\cel{\rlt(2,0)\rlt(0,2)\rlt(-2,0)\rlt(0,-2)\rmt(2,0)}
\def\gcel{\psframe*[linecolor=lightgray](2,2)}
\psset{unit=3.5mm}
\pspicture(-1.5,-2)(14,15)
\put(4,6){\gcel}\put(2,6){\gcel}\put(3,8){\gcel}
\pscustom{
\mt(0,10)\cel\cel\cel\cel\cel\cel
\mt(1,8)\cel\cel\cel\cel\cel
\mt(2,6)\cel\cel\cel\cel
\mt(3,4)\cel\cel\cel
\mt(4,2)\cel\cel
\mt(5,0)\cel
\mt(2.2,4)\rlt(-2,4)\rlt(0,-.5)\rmt(.4,.2)\rlt(-.4,.3)
\mt(4,13)\rlt(4,0)\rlt(-.4,.2)\rmt(.4,-.2)\rlt(-.4,-.2)}
\cput(7,14){$t$}
\cput(.2,5){$w$}
\cput(-1,11){{\footnotesize 6}}
\cput(0,9){{\footnotesize 5}}
\cput(3,3){{\footnotesize 2}}
\cput(4,1){{\footnotesize 1}}
\cput(1,13){{\footnotesize 0}}
\cput(3,13){{\footnotesize 1}}
\cput(9,13){{\footnotesize 4}}
\cput(11,13){{\footnotesize 5}}
\cput(1,11){1}
\cput(3,11){-1}
\cput(5,11){1}
\cput(7,11){-1}
\cput(9,11){1}
\cput(11,11){-1}
\cput(6,9){0}
\cput(2,9){2}
\cput(4,9){-1}
\cput(8,9){1}
\cput(10,9){-2}
\cput(3,7){-2}
\cput(5,7){3}
\cput(7,7){-3}
\cput(9,7){2}
\cput(4,5){-3}\cput(6,5){0}\cput(8,5){3}\cput(5,3){3}\cput(7,3){-3}\cput(6,1){0}
\endpspicture\qquad
\psset{unit=3.5mm}
\pspicture(-1.5,0)(16,17)
\put(8,8){\gcel}\put(6,8){\gcel}\put(7,10){\gcel}
\put(3,6){\gcel}\put(4,4){\gcel}
\pscustom{
\mt(0,12)\cel\cel\cel\cel\cel\cel\cel
\mt(1,10)\cel\cel\cel\cel\cel\cel
\mt(2,8)\cel\cel\cel\cel\cel
\mt(3,6)\cel\cel\cel\cel
\mt(4,4)\cel\cel\cel
\mt(5,2)\cel\cel
\mt(6,0)\cel
\mt(3.3,4)\rlt(-2,4)\rlt(0,-.5)\rmt(.4,.2)\rlt(-.4,.3)
\mt(4,15)\rlt(4,0)\rlt(-.4,.2)\rmt(.4,-.2)\rlt(-.4,-.2)\stroke
}
\cput(7,16){$t$}
\cput(1.2,5){$w$}
\cput(-1,13){{\footnotesize 7}}
\cput(0,11){{\footnotesize 6}}
\cput(1,9){{\footnotesize 5}}
\cput(4,3){{\footnotesize 2}}
\cput(5,1){{\footnotesize 1}}
\cput(1,15){{\footnotesize 0}}
\cput(3,15){{\footnotesize 1}}
\cput(9,15){{\footnotesize 4}}
\cput(11,15){{\footnotesize 5}}
\cput(13,15){{\footnotesize 6}}
\cput(1,13){1}
\cput(3,13){-1}
\cput(5,13){1}
\cput(7,13){-1}
\cput(9,13){1}
\cput(11,13){-1}
\cput(13,13){1}
\cput(6,11){$\frac{1}{2}$}
\cput(8,11){$\frac{1}{2}$}
\cput(2,11) {$\frac{5}{2}$}
\cput(4,11){-$\frac{3}{2}$}
\cput(10,11){-$\frac{3}{2}$}
\cput(12,11){$\frac{5}{2}$}
\cput(3,9){-$\frac{5}{2}$}
\cput(5,9){4}
\cput(7,9){-$\frac{9}{2}$}
\cput(9,9){4}
\cput(11,9){-$\frac{5}{2}$}
\cput(4,7){-$\frac{25}{4}$}\cput(6,7){$\frac{9}{4}$}
\cput(8,7){$\frac{9}{4}$}\cput(10,7){-$\frac{25}{4}$}
\cput(5,5){$\frac{25}{4}$}\cput(7,5){-$\frac{17}{2}$}\cput(9,5){$\frac{25}{4}$}
\cput(6,3){$\frac{17}{4}$}\cput(8,3){$\frac{17}{4}$}
\cput(7,1){-$\frac{17}{4}$}
\endpspicture}
\end{table}

For words with $k-w$ odd, we can use symmetry arguments to find initial values of the~\LOT s, as follows.  Remember that the number of transpositions changes as $t\to w-t-1$ 
under reflection.  For $k$ even, the $A_k$ are anti-symmetric, so that $Z_k(w,t\!-\!1)=-Z_k(w,w\!-\!t)$.  Consequently for even $k$ and odd $w$, $Z_k(w,(w\!-\!1)/2) = 0$.

For $k$ odd, the $A_k$ are symmetric.  Therefore, words that transform into each other under reflection, have the same coefficient: $Z_k(w,t\!-\!1)=Z_k(w,w\!-\!t)$.  
Clearly for even $w$, $Z_k(w,w/2\!-\!1) =  Z_k(w,w/2)$.  Together with the triangle rule (\ref{triangle}) this leads to  $Z_k(w,w/2)= -\frac{1}{2}Z_k(w\!+\!1,w/2) $. 

Continuing with words with $k-w$ even, we noted that in $A_k$ the coefficient of such words with $t = 0$ was always equal to minus the coefficient for words of width $w\!+\!1$ and $t = 0$.  In formula:
\be Z_k(k\!-\!2i,0)=-Z_k(k\!-\!2i\!+\!1,0)\;.\label{k-weven}\ee
Thus for each $w<k$ we now have one value of $t$ for which we know $Z_k(w,t)$. The values for other $t$ can be found with the triangle rule (\ref{triangle}).

This provides a recursive description of all~\LOT s of the coefficients for connected words in $A_k$.   We observed that the lower order terms in these coefficients are equal to the~\LOT~of longer words  as expressed in equation  (\ref{CkinZk}).   Furthermore the coefficients for words with vacancies are the same as the coefficients for connected words with appropriately adjusted values for $w$ and $t$, and  up to an overall factor.  This is explicit  in equation  (\ref{CkwtvginCkwt}).  We emphasize that (\ref{CkwtvginCkwt}) and (\ref{CkinZk}) apply to $A_k$ as well as $Q_k$.

It took considerable effort to discover these regularities in the coefficients of $A_k$.
While we found a recursion to calculate all coefficients, we were unable to find them in closed form.  In appendix \ref{examplesQk} we give the first seven elements of the series $A_k$.   Later we discovered that not all of the observed properties are essential for  $[A_k, H]=0$.  In particular, the symmetry of $A_k$ is constructed, and also (\ref{k-weven}) turns out to be unnecessary.  It is this freedom that allowed us to come up with the closed form (\ref{eq:Z-table}).

\section{Proof}\label{SecProof}
We begin with writing the commutator analogous to (\ref{QkinDkq})
\be [Q_k,H] = \sum_{p} S_k(p)\,p \label{QkHinSkp} \ee
where the sum is over all unique irreducible words in the algebra.   
We will prove that for $Q_k$ given by equations (\ref{CkwtvginCkwt}-\ref{QkinCk}),  $S_k(p) = 0$ for any $p$.   

In general, $ S_k(p)$ can be calculated as
\be S_k(p) = \sum_{q,\sg\,|\,p\,=\,q\,e_\sg} D_k(q)\; -  \sum_{q,\sg\,|\,p\,=\,e_\sg\,q} D_k(q) \label{RkinDk}\ee
summed over all $q \in {\rm TL_1}$ and indices $\sg\in \Z_L$ such that the product of $q$ and $e_\sg$ equals  $p$.

Since we already know that $Q_k$ has only words with single occurrence of the monoids, the reduced words in $[Q_k,H]$, can have at most one monoid that appears twice, and no monoid appearing more often.
Let $p$ be a reduced word in which  (only) $e_\sg$ appears twice.    Then the sums in (\ref{RkinDk}) reduce to one term:
\be S_k(p) = D_k(q) - D_k(q')\,, \quad\mbox{where}\quad p = q \, e_\sg = e_\sg\,q' \ee
and the word $q$ contains the sequence
$... e_\sg \,e_{\sg\!-\!1}...e_{\sg\!+\!1}...$ whereas $q'$ contains
$...  e_{\sg\!-\!1}...e_{\sg\!+\!1}\, e_\sg...$.   The words $q$ and $q'$ agree in all details except the order of $e_\sg$ and $e_{\sg\pm1}$.    
In both $q$ and $q'$ the order of $e_\sg$ and $e_{\sg\pm1}$ constitutes one transposition, and the distribution of vacancies is identical.   Therefore, with (\ref{DkinCkwtvg}), $D_k(q) = D_k(q')$ and consequently $S_k(p)=0$ for all irreducible words $p$ in which a monoid appears twice.

What remains is the words $p\in{\rm TL_1}$.
Figure~\ref{wordinTL1} can be used to illustrate some of the relevant properties.
If we assume that a word $p$ appears in $[Q_k, H]$, it must be the reduced form of $q\,e_\sg$ or of $e_\sg\,q$, where $q\in{\rm TL_1}$.
Before reduction, the position of  $e_\sg$ in $p$ is initial or final.
Since the rules (\ref{TL-rules})  cannot create or annihilate an initial or final monoid,
the monoid $e_\sg$ is still present in the reduced form and is still initial or final.

In the word in figure~\ref{wordinTL1} the monoids $e_1$, $e_4$, $e_6$, $e_{12}$,  $e_{17}$ and $e_{19}$ are initial, i.e.  can play the role of $e_\sg$ in  $p=q\,e_\sg$.  Likewise the monoids $e_2$, $e_{10}$, $e_{13}$,  $e_{17}$ and $e_{20}$ are final, and can thus be  $e_\sg$ in $p=e_\sg\,q$.
We say that these monoids 'contribute' $\pm D_k(q)$ to $S_k(p)$, see (\ref{RkinDk}).
What these contributions can be, depends on the direct environment of $e_\sg$ in the word $p$.   We will label these different environments and the corresponding contribution to $S_k(p)$ with a two-letter code as shown at the top of figure \ref{wordinTL1}.
The code for a monoid $e_\sg$ has a letter $\P$ for each neighbor it is preceding, and a letter $\F$ for each neighbor it is following.  
Thus $e_1$, $e_4$, $e_6$ and $e_{19}$ have one $\P$, and $e_{12}$ has two $\P$'s.
The codes PF or FP we ignore, because the corresponding monoid must be positioned between its neighbors.  Its contribution is zero, because it is neither initial nor final.

Each monoid which has a vacancy on its left (right) carries an $\L$ ($\R$) in its code.
The $\L$ and $\R$ have a suffix, 1 in case the neighboring vacancy forms a one-vacancy gap, 2 if it is part of a larger gap or 3 if it marks (left or right) end of the word.  
Thus, each contributing monoid has two symbols in its code, a $\P$ or an $\F$ for each neighboring monoid in the word, or an $\L$ or an $\R$ for a neighboring vacancy.   

\begin{table}[t]\caption[Contributions to $S_k(p)$ of a monoid preceding its neighbors]{List of the contributions of the monoid $e_\sg$ in a word $p$ to its coefficient $S_k(p)$ in $[Q_k,H]$.   In this table  $e_\sg$ is {\em preceding} one of its neighbors.   The word $p$ has width $w$, and contains $t$ transpositions, $v$ vacancies and $g$ gaps.
Column 1 has the code for  the environment of $e_\sg$ in $p$, column 3 shows the possible configurations of monoids in $e_\sg\;q$ or  $q\,e_\sg$ before reduction, that give rise to the coded environment in the reduced version of $p$.   In column 3 we have $q$ in parentheses, to distinguish it from $e_\sg$.  Column 2 shows the range of monoids of which the presence and positions are specified in the third column.   Column 4 gives the resulting contribution to $S_k(p)$.  }
\label{tableP}
\[  \begin{array}{cccl}
\mbox{code}&\mbox{specified}&\mbox{configuration}&\mbox{contribution}\\
\hline
{\rm PL_1}&\rng{\sg\!-\!2}{\sg\!+\!1}
	&(...e_{\sg-2}...e_{\sg+1}\,e_{\sg}\,e_{\sg-1}...)\,e_{\sg}	&C_k(w,t\!+\!1,v\!-\!1,g\!-\!1)\\
% e_{\sg-2}...e_{\sg+1}...e_{\sg}
&	&(...e_{\sg+1}\,e_{\sg}\,e_{\sg-1}...e_{\sg-2}...)\,e_{\sg}	&C_k(w,t\!+\!2,v\!-\!1,g\!-\!1)\\
&	&(...e_{\sg-2}...e_{\sg+1}\,e_{\sg}...)\,e_{\sg}		&\tau\,C_k(w,t,v,g)\\
&	&(...e_{\sg-2}...  e_{\sg+1}...)\,e_{\sg}			&C_k(w,t\!-\!1,v\!+\!1,g)\\
\hline
{\rm PR_1}&\rng{\sg\!-\!1}{\sg\!+\!2}
	&(...e_{\sg-1}...e_{\sg}\,e_{\sg+1}...e_{\sg+2}...)\,e_{\sg}	&C_k(w,t,v\!-\!1,g\!-\!1)\\
&	&(...e_{\sg-1}...e_{\sg}...e_{\sg+2}\,e_{\sg+1}...)\,e_{\sg}	&C_k(w,t\!+\!1,v\!-\!1,g\!-\!1)\\ 
&	&(...e_{\sg-1}...  e_{\sg}...e_{\sg+2}...)\,e_{\sg}		&\tau\,C_k(w,t,v,g)\\
&	&(...  e_{\sg-1} ...e_{\sg+2}...)\,e_{\sg}			&C_k(w,t,v\!+\!1,g)\\
\hline
{\rm PL_2}&	\rng{\sg\!-\!2}{\sg\!+\!1}
	&(...e_{\sg+1}...e_{\sg}\,e_{\sg-1}...)\,e_{\sg}		&C_k(w,t\!+\!1,v\!-\!1,g)\\
&	&(...e_{\sg+1}...e_{\sg}...) \,e_{\sg}				&\tau\,C_k(w,t,v,g)\\
&	&(...e_{\sg+1}...) \,e_{\sg}					&C_k(w,t\!-\!1,v\!+\!1,g)\\
\hline
{\rm PR_2}&	\rng{\sg\!-\!1}{\sg\!+\!2}
	&(...e_{\sg-1}...e_{\sg}\,e_{\sg+1})\,e_{\sg}			&C_k(w,t,v\!-\!1,g)\\
 &	&(...  e_{\sg-1}...e_{\sg}) \,e_{\sg}				&\tau\,C_k(w,t,v,g)\\
&	&(...e_{\sg-1}...) \,e_{\sg}					&C_k(w,t,v\!+\!1,g)\\
\hline
{\rm PL_3}&\rng{\sg\!-\!1}{\sg\!+\!1}
	&(...e_{\sg+1}...e_{\sg}\,e_{\sg-1}...)\,e_{\sg}		&C_k(w\!+\!1,t\!+\!1,v,g)\\
&	&(...e_{\sg+1}...e_{\sg}...) \,e_{\sg}				&\tau\,C_k(w,t,v,g)\\
&	&(...e_{\sg+1}...) \,e_{\sg}					&C_k(w\!-\!1,t\!-\!1,v,g)\\     
\hline
{\rm PR_3}&	\rng{\sg\!-\!1}{\sg\!+\!1}
	&(...e_{\sg-1}...e_{\sg}\,e_{\sg+1})\,e_{\sg}			&C_k(w\!+\!1,t,v,g)\\
&	&(...  e_{\sg-1}...e_{\sg}) \,e_{\sg}				&\tau\,C_k(w,t,v,g)\\
&	&(...e_{\sg-1}...) \,e_{\sg}					&C_k(w\!-\!1,t,v,g)\\
\hline
{\rm PP}&	\rng{\sg\!-\!1}{\sg\!+\!1}
	&(...e_{\sg-1}...e_{\sg+1}...e_\sg...)\,e_\sg			&\tau\,C_k(w,t,v,g)\\
&	&(...e_{\sg-1}...e_{\sg+1}...) \,e_\sg				& C_k(w,t\!-\!1,v\!+\!1,g+1)\\
\hline\end{array}\]\end{table}
\begin{figure}[b]
\def\mnd{\pscircle*[linecolor=\choice](0.5,0.5){.69}
\pscustom[linecolor=blue,linewidth=1.5pt]{
\mt(0,0) \rct(0,.4)(1,.4)(1,0)\mt(0,1) \rct(0,-.4)(1,-.4)(1,0)}}
\def\choice{pink}
\psset{unit=4mm}
\def\spcl{\def\choice{lgr}\md{3,-1}}
\def\spc{\hspace*{2cm}}
\spc\pspicture(0,-3)(4,5)\def\vln{\rlt(0,5)\rmt(1,-5)}
\pscustom [linecolor=blue,linewidth=1.5pt]{
\mt(2,0)\vln\vln\vln}
\md{1,2}\md{4,3}\def\choice{ylw}\md{3,2}
\put(-1,-3){$S_k(w,t,v,g) = \cdots$}
\endpspicture
\pspicture(-5.5,-3)(5,7)\def\vln{\rlt(0,6)\rmt(1,-6)}
\pscustom [linecolor=blue,linewidth=1.5pt]{
\mt(2,0)\vln\vln\vln}
\md{1,3}\md{2,2}\md{3,3}\md{4,4}\spcl
\put(-2,-3){$+  \;C_k(w,t\!+\!1,v\!-\!1,g\!-\!1)$}
\endpspicture
\pspicture(-7.5,-3)(5,7)\def\vln{\rlt(0,6)\rmt(1,-6)}
\pscustom [linecolor=blue,linewidth=1.5pt]{
\mt(2,0)\vln\vln\vln}
\md{1,1}\md{2,2}\md{3,3}\md{4,4}\spcl
\put(-3,-3){$+\;\; C_k(w,t\!+\!2,v\!-\!1,g\!-\!1)$}
\endpspicture
\\
\spc\pspicture(-10,-3)(5,7)\def\vln{\rlt(0,7)\rmt(1,-7)}
\pscustom [linecolor=blue,linewidth=1.5pt]{
\mt(2,-1)\vln\vln\vln}
\md{1,2}\md{3,2}\md{4,3}\spcl
\put(-2,-3){$+\;\;\tau\, C_k(w,t,v,g) $}
\endpspicture
\pspicture(-5.5,-3)(5,5)\def\vln{\rlt(0,6)\rmt(1,-6)}
\pscustom [linecolor=blue,linewidth=1.5pt]{
\mt(2,-1)\vln\vln\vln}
\md{1,2}\md{4,2}\spcl
\put(-3,-3){$+\quad C_k(w,t\!-\!1,v\!+\!1,g) \; \;\cdots $}
\endpspicture
\caption[Picture of environment PL$_1$]{\label{fig:PL1} The contributions to a word $p$ in $[Q_k,H]$ from a monoid $e_\sg$ which has a one-vacancy gap on its left, and is preceding its right neighbor, corresponding to the code $\P\L_1$.  The leftmost diagram in the top line is the environment of $\sg$ in $p$, and the other diagrams are the environments in the possible words $q$ in $Q_k$, which contribute to the coefficient of $p$, as in equation (\ref{RkinDk}).  The contributions to $S_k(w,t,v,g)$ corresponding to the diagrams are written under them.  The figures and the expressions are explained in the text. }
\end{figure}

In the tables \ref{tableP}-\ref{tableS}, we list for each environment of $e_\sg$ in a word $p$ its contribution to $S_k(p)$ according to (\ref{RkinDk}).   To avoid unwieldy tables, the contributions are divided in three categories: table \ref{tableP} for monoids which are {\em preceding} at least one of their neighbors, table \ref{tableF} for monoids which are {\em following} one or both of their neighbors and table \ref{tableS} for monoids of which no neighbor is present in the word.   

To show how to construct (\& read) the tables we discuss the first code in  table \ref{tableP},  $\P\L_1$, as illustrated in figure \ref{fig:PL1}.
This code implies that the word $p$ in $[Q_k,H]$ contains 
$...e_{\sg-2}...e_{\sg+1}\,e_{\sg}...$ where only the monoids in the range $[\sg\!-\!2,\sg\!+\!1]$ (column 2) are specified. The string in the third column implies that  $e_{\sg-1}$ is absent, leaving a single vacancy, and $e_{\sg}$ precedes $e_{\sg+1}$.   Other monoids can be placed in the positions of the dots, where we have used the convention that indices of the monoids are increasing unless prevented by non-commutation.   With this code $\P\L_1$, $p$ can only be the reduced form of $q\,e_\sg$ (and not of $e_\sg\,q$), where $q$ is a word in $Q_k$.

The possibilities for $q$ are specified in the third column.  
In the first case, $q=...e_{\sg-2}...e_{\sg+1}\,e_{\sg}\,e_{\sg-1}...$, the word $p=q\,e_\sg$ is reduced by $e_{\sg}\,e_{\sg-1}\,e_{\sg}=e_{\sg}$, (\ref{TL-rules}), creating the vacancy at position $\sg-1$.   
The word $q$ has one transposition more than $p$ (the sequence $e_{\sg}\,e_{\sg-1}$), one vacancy less, as $e_{\sg-1}$ is present in $q$, and also one gap less.   Thus its contribution to $S_k(p)$ is equal to: $C_k(w,t\!+\!1,v\!-\!1,g\!-\!1)$, shown in column four.

The next line in the table has the word $...e_{\sg+1}\,e_{\sg}\,e_{\sg-1}...e_{\sg-2}...$; very similar, but now with $e_{\sg-1}$ following $e_{\sg-2}$ rather than preceding it.
%But the end result for $p$ is the same, as $e_\sg$   and  $e_{\sg-2}$ commute.   
Now $q$ has two more transpositions than $p$, hence the contribution  $C_k(w,t\!+\!2,v\!-\!1,g\!-\!1)$.

The third line shows the case that $q=...e_{\sg-2}...e_{\sg+1}\,e_{\sg}...$, i.e.  equal to $p$.   Appending it with $e_\sg$ does not change the word, only its coefficient, as $e_\sg^2=\tau\,e_\sg$.   The resulting contribution is thus $\tau\,C_k(w,t,v,g)$.  Finally, it is also possible that $e_\sg$ is not contained in $q$, which then reads $...e_{\sg-2}...  e_{\sg+1}...$.  In this case the single vacancy in $p$ at position $\sg-1$ is part of a two-vacancy gap  in $q$, thus not changing the number of gaps.   And the transposition of $e_{\sg+1}\,e_{\sg}$ is gone, so that the contribution to $S_k(p)$ is $C_k(w,t\!-\!1,v\!+\!1,g)$.   
The other lines in the table can be read in the same way.

Table \ref{tableF} has the same structure.   In the ordering of the lines we have interchanged the code R and L, relative to table \ref{tableP}.   This makes it manifest that the contributions in table \ref{tableP} are precisely the negative of those in  \ref{tableF}, row by row.
\begin{table}[bt]\caption[Contributions to $S_k(p)$ of a monoid following its neighbors]{List of the contributions of the monoid $e_\sg$ in a word $p$ to its coefficient $S_k(p)$ in $[Q_k,H]$.   In this table  $e_\sg$ is {\em following} one or both of its neighbors, hence the $\F$ in the code.  }\label{tableF}
\[  \begin{array}{cccl}
\mbox{code}&\mbox{specified}&\mbox{configuration}&\mbox{contribution}\\
\hline
{\rm FR_1}&\rng{\sg\!-\!1}{\sg\!+\!2}
	&e_{\sg}\,(...e_{\sg+1}\,e_{\sg}\,e_{\sg-1}...e_{\sg+2}...)	&-C_k(w,t\!+\!1,v\!-\!1,g\!-\!1)\\
&	&e_{\sg}\,(...e_{\sg+2}\,e_{\sg+1}\,e_{\sg}\,e_{\sg-1}...)	&-C_k(w,t\!+\!2,v\!-\!1,g\!-\!1)\\ 
&	&e_{\sg}\,(...e_{\sg}\,e_{\sg-1}...e_{\sg+2}...)                      &-\tau\,C_k(w,t,v,g)\\
&	&e_{\sg}\,(...e_{\sg-1}...e_{\sg+2}...)                                 &-C_k(w,t\!-\!1,v\!+\!1,g)\\            
\hline
{\rm FL_1}&\rng{\sg\!-\!2}{\sg+1}
	&e_{\sg}\,(...e_{\sg-2}...e_{\sg-1}\,e_{\sg}...e_{\sg+1}...)	&-C_k(w,t,v\!-\!1,g\!-\!1)\\
&	&e_{\sg}\,(...e_{\sg-1}\,e_{\sg-2}\,e_{\sg}...  e_{\sg+1}...)	&-C_k(w,t\!+\!1,v\!-\!1,g\!-\!1)\\
&	&e_{\sg}\,(...e_{\sg-2}...e_{\sg}...  e_{\sg+1}...)		&-\tau\,C_k(w,t,v,g)\\
&	&e_{\sg}\,(...e_{\sg-2}...  e_{\sg+1}...)			&-C_k(w,t,v\!+\!1,g)\\
\hline
{\rm FR_2}&\rng{\sg\!-\!1}{\sg\!+\!2}
	&e_{\sg}\,(...e_{\sg+1}\, e_{\sg}...  e_{\sg-1}...)		&-C_k(w,t\!+\!1,v\!-\!1,g)\\
&	&e_{\sg}\,(...e_{\sg}...  e_{\sg-1}...)				&-\tau\,C_k(w,t,v,g)\\
&	&e_{\sg}\,(...  e_{\sg-1}...)					&-C_k(w,t\!-\!1,v\!+\!1,g)\\
\hline
{\rm FL_2}&	\rng{\sg\!-\!2}{\sg+1}
	&e_{\sg}\,(e_{\sg-1}\, e_{\sg}...  e_{\sg+1}...)			&-C_k(w,t,v\!-\!1,g)\\
&	&e_{\sg}\,(e_{\sg}...  e_{\sg+1}...)				&-\tau\,C_k(w,t,v,g)\\
&	&e_{\sg}\,(...  e_{\sg+1}...)					&-C_k(w,t,v\!+\!1,g)\\
\hline
{\rm FR_3}&	\rng{\sg\!-\!1}{\sg\!+\!1}
	&e_{\sg}\,(...e_{\sg+1}\, e_{\sg}...  e_{\sg-1}...)		&-C_k(w\!+\!1,t\!+\!1,v,g)\\
&	&e_{\sg}\,(...e_{\sg}...  e_{\sg-1}...)				&-\tau\,C_k(w,t,v,g)\\
&	&e_{\sg}\,(...  e_{\sg-1}...)					&-C_k(w\!-\!1,t\!-\!1,v,g)\\
\hline
{\rm FL_3}&\rng{\sg\!-\!1}{\sg\!+\!1}
	&e_{\sg}\,(e_{\sg-1}\, e_{\sg}...  e_{\sg+1}...)			&-C_k(w\!+\!1,t,v,g)\\
&	&e_{\sg}\,(e_{\sg}...  e_{\sg+1}...)				&-\tau\,C_k(w,t,v,g)\\
&	&e_{\sg}\,(...  e_{\sg+1}...)					&-C_k(w\!-\!1,t,v,g)\\
\hline
{\rm FF}&	\rng{\sg\!-\!1}{\sg\!+\!1}
	&e_\sg\,(...e_\sg...e_{\sg-1}...e_{\sg+1}...)			&-\tau\,C_k(w,t,v,g)\\
&	&e_\sg\,(...e_{\sg-1}...e_{\sg+1}...)				&- C_k(w,t\!-\!1,v\!+\!1,g\!+\!1)\\
\hline\end{array}\]\end{table} 

A separate table shows the contributions for an isolated  monoid between two vacancies, (e.g.  $e_{17}$ in figure~\ref{wordinTL1}).   It can have nine different environments (on both sides a 1, 2 or 3).   An isolated monoid could simply be created by an $e_\sg$ in $H$, inside a larger gap, or contracted with an existing isolated  $e_\sg$ by $e_\sg^2 = \tau\,e_\sg$.   However, these terms cancel in $[q,e_\sg ]$, so these events do not contribute, and we do not list them.   Non-cancelling terms results from the events that in $q$ the monoid $e_\sg$ is preceded or followed by one of its neighbors $e_{\sg\pm1}$, and that this neighbor is annihilated by the multiplication by $e_\sg$ from $H$.  In the interest of brevity, we separate the terms associated with the left and right neighbor, and label them accordingly with $\L$ and $\R$.
The total contribution for an isolated monoid is the sum of its contributions labelled as $\L_i$ and $\R_j$, and with this sum we associate the code $\L_i\R_j$.  
 \begin{table}[th]\caption[Contribution to $S_k(p)$ of an isolated monoid]{Contributions of an isolated monoid in between two vacancies.    Its code specifies only the nature of the left and right vacancies.   The contributions involving annihilation of the monoids on the left and on the right are separated.   Each $\L_i$ should be combined with an $\R_j$, written as $\L_i\R_j$, nine possibilities in total..  }\label{tableS}
\[\begin{array}{cccl}	
\mbox{code} & \mbox{range} & \mbox{configuration} & \mbox{contribution} \\
\hline 
{\rm L_1}&\rng{\sg\!-\!2}{\sg\!+\!1}
		& e_{\sg}\,(e_{\sg-2}\,e_{\sg-1}\,e_{\sg})      &-C_k(w,t,v\!-\!1,g\!-\!1)      \\
	&	& e_{\sg}\,(e_{\sg-1}\,e_{\sg-2}\,e_{\sg})      &-C_k(w,t\!+\!1,v\!-\!1,g\!-\!1)    \\
	&	& (e_{\sg-2}\,e_{\sg}\,e_{\sg-1})\,e_{\sg}      &+C_k(w,t\!+\!1,v\!-\!1,g\!-\!1)    \\
	&	& (e_{\sg}\,e_{\sg-1}\,e_{\sg-2})\,e_{\sg}      &+C_k(w,t\!+\!2,v\!-\!1,g\!-\!1)    \\
\hline
{\rm L_2}&\rng{\sg\!-\!2}{\sg\!+\!1}
		& e_{\sg}\,(e_{\sg-1}\,e_{\sg})	&-C_k(w,t,v\!-\!1,g)		\\
	&	& (e_{\sg}\,e_{\sg-1})\,e_{\sg}	&+C_k(w,t\!+\!1,v\!-\!1,g)		\\
\hline
{\rm L_3}&\rng{\sg\!-\!2}{\sg\!+\!1}
		& e_{\sg}\,(e_{\sg-1}\,e_{\sg})	    	&-C_k(w\!+\!1,t,v,g)	\\
	&	& (e_{\sg}\,e_{\sg-1})\,e_{\sg}	    	&+C_k(w\!+\!1,t\!+\!1,v,g)	\\
\hline
{\rm R_1}&\rng{\sg\!-\!1}{\sg\!+\!2}
		& (e_{\sg}\,e_{\sg+1}\,e_{\sg+2})\,e_{\sg}	&+C_k(w,t,v\!-\!1,g\!-\!1)		\\
	&	& (e_{\sg}\,e_{\sg+2}\,e_{\sg+1})\,e_{\sg}	&+C_k(w,t\!+\!1,v\!-\!1,g\!-\!1) 	\\
	&	& e_{\sg}\,(e_{\sg+1}\,e_{\sg}\,e_{\sg+2}) 	&-C_k(w,t\!+\!1,v\!-\!1,g\!-\!1)	\\
	&	& e_{\sg}\,(e_{\sg+2}\,e_{\sg+1}\,e_{\sg}) 	&-C_k(w,t\!+\!2,v\!-\!1,g\!-\!1)	\\
\hline
{\rm R_2}&\rng{\sg\!-\!1}{\sg\!+\!2}
		& (e_{\sg}\,e_{\sg+1})\,e_{\sg}	&+C_k(w,t,v\!-\!1,g)		\\
	&	& e_{\sg}\,(e_{\sg+1}\,e_{\sg}) 	&-C_k(w,t\!+\!1,v\!-\!1,g)		\\
\hline
{\rm R_3}&\rng{\sg\!-\!1}{\sg\!+\!2}
		& (e_{\sg}\,e_{\sg+1})\,e_{\sg}	    	&+C_k(w\!+\!1,t,v,g)	\\
	&	& e_{\sg}\,(e_{\sg+1}\,e_{\sg}) 	    	&-C_k(w\!+\!1,t\!+\!1,v,g)	\\
\hline
\end{array}\]\end{table}

The contributions to $S_k(p)$ due to an isolated monoid in $p$ positioned between a single-vacancy gap on its right, and a larger gap on its left, the environment $\L_2\R_1$, are shown in figure~\ref{fig:L2R1}. The first two contributions come from the event that the gap on the left is enlarged, as result of multiplication by a monoid from $H$, listed with code $\L_2$ in table \ref{tableS}.  The other four contributions result from the event that the one-vacancy gap is created by a $e_\sg$ from $H$, listed with code $R_1$.
\begin{figure}[ht]
  \caption[Picture of environment $\L_2\R_1$]{\label{fig:L2R1} The contributions to $S_k(p)$ from a monoid with environment code $\L_2\R_1$.  The monoid $e_\sg$ from $H$ is shown at the top (from $H\,Q_k$) or bottom (from $Q_k\,H$) of each diagram.  
The first two diagrams show the creation of a vacancy at position $\sg\!-\!1$, and the other four at position $\sg\!+\!1$.}
\def\mnd{\pscircle*[linecolor=\choice](0.5,0.5){.69}
\pscustom[linecolor=blue,linewidth=1.5pt]{
\mt(0,0) \rct(0,.4)(1,.4)(1,0)\mt(0,1) \rct(0,-.4)(1,-.4)(1,0)}}
\def\choice{pink}
\psset{unit=4mm}\qquad
\pspicture(0,-1)(5,7)\def\vln{\rlt(0,5)\rmt(1,-5)}
\pscustom [linecolor=blue,linewidth=1.5pt]{
\mt(1,0)\vln\vln\vln\vln}
\md{4,2}\def\choice{ylw}\md{2,2}
\cput(4,-2){$S_k(w,t,v,g) =\ldots$}
\endpspicture\qquad\qquad
\pspicture(0,-.5)(5,6.5)\def\vln{\rlt(0,5)\rmt(1,-5)}
\pscustom [linecolor=blue,linewidth=1.5pt]{
\mt(1,0)\vln\vln\vln\vln}
\md{4,1}\md{2,1}\md{1,2}\def\choice{lgr}\md{2,5}
\cput(3,-1.5){\small$-\;C_k(w,t,v\!-\!1,g) $}
\endpspicture\qquad\qquad
\pspicture(0,-1.5)(6,4)\def\vln{\rlt(0,5)\rmt(1,-5)}
\pscustom [linecolor=blue,linewidth=1.5pt]{
\mt(1,0)\vln\vln\vln\vln}
\md{4,3}\md{2,3}\md{1,2}\def\choice{lgr}\md{2,-1}
\cput(3.2,-2.5){\small$+\;\;C_k(w,t\!+\!1,v\!-\!1,g) $}
\endpspicture\qquad\qquad 
\pspicture(-1,-.5)(5,6.5)\def\vln{\rlt(0,5)\rmt(1,-5)}
\pscustom [linecolor=blue,linewidth=1.5pt]{
\mt(1,0)\vln\vln\vln\vln}
\md{4,1}\md{2,1}\md{3,2}\def\choice{lgr}\md{2,5}
\cput(3.5,-1.5){\small$-\;C_k(w,t\!+\!1,v\!-\!1,g\!-\!1) $}
\endpspicture\\
\pspicture(-10,-4)(7,8)\def\vln{\rlt(0,5)\rmt(1,-5)}
\pscustom [linecolor=blue,linewidth=1.5pt]{
\mt(1,0)\vln\vln\vln\vln}
\md{4,3}\md{2,3}\md{3,2}\def\choice{lgr}\md{2,-1}
\cput(2,-2){\small$+\;C_k(w,t\!+\!1,v\!-\!1,g\!-\!1) $}
\endpspicture
\pspicture(-3,-4)(7,8)\def\vln{\rlt(0,5)\rmt(1,-5)}
\pscustom [linecolor=blue,linewidth=1.5pt]{
\mt(1,0)\vln\vln\vln\vln}
\md{4,3}\md{2,1}\md{3,2}\def\choice{lgr}\md{2,5}
\cput(2,-2){\small$-\;C_k(w,t\!+\!2,v\!-\!1,g\!-\!1) $}
\endpspicture
\pspicture(-2.5,-4)(7,8)\def\vln{\rlt(0,5)\rmt(1,-5)}
\pscustom [linecolor=blue,linewidth=1.5pt]{
\mt(1,0)\vln\vln\vln\vln}
\md{4,1}\md{2,3}\md{3,2}\def\choice{lgr}\md{2,-1}
\cput(3,-2){\small$+\;C_k(w,t,v\!-\!1,g\!-\!1) \;\cdots$}
\put(4,-2){\ldots}
\endpspicture  
 \end{figure}

    When for a word $p$ all the single monoid contributions add up to zero, this proves that $S_k(p) = 0$.   We will set out to prove that  $S_k(p) = 0$ for all $p$, and hence $[Q_k,H]=0$.
The number of environments of a single monoid in a word giving rise to distinct contributions is 23, as listed in the tables \ref{tableP}, \ref{tableF} and \ref{tableS}.
It is clear by inspection that under left-right reflection,  i.e.  $\F\leftrightarrow\P$ and $\L\leftrightarrow\R$, the contributions change sign.   In table \ref{tableP} and \ref{tableF} the rows are ordered such that they manifestly cancel line by line, irrespective of form of $C_k(w,t,v,g)$.   In table \ref{tableS} the first three cases are the negatives of the last three, so that the contributions of  $\L_i\R_j$ form an antisymmetric matrix.   This reduces the 23 distinct contributions of single monoids to ten independent ones, say the seven codes of table \ref{tableP}, and the codes $\L_1\R_2$, $\L_2\R_3$, $\L_3\R_1$.
%This fact does not depend on the form of $C_k(w,t,v,g)$, but only on the property that it is fully determined by the four arguments $\{w,t,v,g\}$.

%We will proceed to show that all the contributions of a single monoid is the sum of two parts corresponding to the two components of its label.   So there is a contribution for the $\P$ and an opposite contribution for the $\F$.   Likewise there is a contribution for the $\L_1$, $\L_2$ and $\L_3$, and the opposite for  $\R_1$, $\R_2$ and $\R_3$.   Then it is not difficult to prove that the contribution for an entire word vanishes.  Clearly a one-vacancy gap has an $\L_1$ (an $\R_1$) in the label of the monoid on its right (left) side.   Analogously a larger gap has a $\L_2$ and $\R_2$.   And the whole word has one $\L_3$ and one $\R_3$.   Similarly each connected sequence of monoids in the word with uniformly increasing indices, $e_\mu\, e_{\mu+1}\, e_{\mu+2}...$ or uniformly decreasing indices has an $\F$ in the code of the last monoid and a $\P$ in the code of the first.   Clearly for every $\P$ there is an $\F$.   It remains to prove that the contributions of single monoids is additive in the two components of their labels.   

From here on our strategy is as follows: we will prove that the remaining ten independent contributions satisfy six independent equations.  Then we will give a four-parameter solution for all contributions.  Since the equations are linear, it is straightforward to check their independence (the rank of the coefficient matrix is equal to six).  For the same reason  the completeness of the solution is guaranteed.  Finally, we will make clear that the structure of solution guarantees that for all words $p$ in $[H, Q_k]$, the total of all single monoid-contributions, i.e. the coefficient $C_k(p)$, vanishes.

We will first list the six simple linear equations in terms of the single-monoid contributions, denoted with the code of the corresponding environment.
\begin{eqnarray}
{\rm L_3R_1 + L_1R_2 + L_2R_3 } & = & 0 \label{LRLRLR} \\   %  0 ; 0  
{\rm FL_3 + PR_2 + L_2R_3 } & = & 0 \label{FLPRLR} \\         % 10 ; 0
{\rm PL_3 + FR_2 + L_2R_3 } & = & 0 \label{PLFR2L2R} \\      % 11 ; 0
{\rm PL_3 + FR_1 + L_1R_3 } & = & 0 \label{PLFR1L1R} \\      %  9 ; 1 
{\rm PL_3 + FF + PR_3 } & = & 0 \label{PL3FFPR3} \\            %  7 ; 2
{\rm PL_1 + FF + PR_1 } & = & 0 \label{PL1FFPR1}             %  5 ; 4
\end{eqnarray}

Before we proceed to prove these equations, a brief comment on their structure.  These equations form a somewhat arbitrary selection out of many (empirical) possibilities.  The criteria for the choice are (i) independence in terms of the ten independent contributions mentioned above, (ii) simplicity, and (iii) a total number of six.
The left-hand side of equation  (\ref{LRLRLR}) can be interpreted as the total coefficient $S_k(p)$ for a word $p$ which consist of three single monoids, of which the left-most is separated from the middle one by one vacancy, while the middle one is more distant from the right-most monoid.   But it can also be the contribution to  $S_k(p)$ of three isolated monoids with the named properties, in a word consisting of more monoids.    Similarly the Eqs.  (\ref{FLPRLR})-(\ref{PL3FFPR3}) can be interpreted as the total $S_k(p)$ of an entire word $p$.   In the graphical representation shown in  figure~\ref{wordinTL1}, the LHS of (\ref{FLPRLR}), for instance, is the coefficient of a word consisting of a descending sequence of monoids on the left end of the word,  separated by  a gap of more than one vacancy from a single monoid at the right end.
Equation  (\ref{PL1FFPR1}) on the other hand does not include an index 3, so it cannot be the total coefficient of a word.   It is instead, the contribution of a connected string between two single-vacancy gaps, consisting of an ascending sequence of monoids on the left, and a descending one on the right.  

We will now proceed to prove the equations (\ref{LRLRLR})-(\ref{PL3FFPR3}) successively.
\begin{itemize}
\item
[Eq.  (\ref{LRLRLR})] is obviously valid, as it is clear from table \ref{tableS} that the contributions coded $\L_i\R_j$ can be written as  $\L_i\R_j=\Lambda_i-\Lambda_j$.  \hfill$\Box$
\item
[Eqs. (\ref{FLPRLR}, \ref{PLFR2L2R})]
When in these equations the contributions from the tables are substituted for their codes, the resulting equations, expressed in $C_k(w,t,v,g)$ are not manifestly true; but expressed in $C_k(w,v)$ via (\ref{CkwtvginCkwt}) they are.  \hfill$\Box$
\item
[Eq.  (\ref{PLFR1L1R})] turns out to be slightly more involved.   Expressing the equation in $C_k(w,t)$ results in
\be -(-\tau)^{g-1} \,\Delta_k(W\!-\!2,T\!-\!1) = 0 ,\label{DeltaZero}
\ee
where we have introduced the shorthand
\be W := w\!+\!v\!+\!g \qquad\text{ and }\qquad T:= t\!+\!v\!+\!g\;.\ee
and the function
\be \Delta_k(W,T) \;:=\; C_k(W,T\!-\!1)\;+\;C_k(W,T)\;+\;\tau\,C_k(W\!+\!1,T)\,.\ee 
To verify the validity of (\ref{DeltaZero}) we express
$\Delta_k$ in terms of $Z_k$ using (\ref{CkinZk}) 
\[
\Delta_k(w,t) =  \sum_{j=0}^{(k-w)/2}\tau^{k-w-2j} \; \big[Z_k(w\!+\!2j,\,t\!+\!j\!-\!1) + Z_k(w\!+\!2j,\,t\!+\!j)\big] \;+ \qquad \] \be \qquad
\sum_{j=0}^{(k-w\!-\!1)/2}\tau^{k-w-2j}\; Z_k(w\!+\!2j\!+\!1,\,t\!+\!j) \ee
Since $Z_k(w,t)$ vanishes for $w>k$ , one can freely extend the last summation with one term (if needed) by giving it the same upper limit as the first, so that
\be
\Delta_k(w,t) = \!\! \sum_{j=0}^{(k-w)/2}\tau^{k-w-2j}  \big[Z_k(w\!+\!2j,\,t\!+\!j\!-\!1) + Z_k(w\!+\!2j,\,t\!+\!j)
+Z_k(w\!+\!2j\!+\!1,\,t\!+\!j) \big] \ee
From (\ref{triangle}) it follows that \be \Delta_k(w,t)=0\qquad\text{ if }\qquad 0<t<w ,\label{Delta}\ee so the only question remaining, is if the arguments of $\Delta_k$ in (\ref{DeltaZero}) satisfy the condition of (\ref{Delta}): $0<T\!-\!1<W\!-\!2$  as (\ref{PLFR1L1R}) is applied to a possible word in $[Q_k,H]$.   Such a word must have at least  one gap of one vacancy, which implies that $v\!+\!g \geq 2$; and it has at least one ascending sequence, so that indeed $T = t\!+\!v\!+\!g > 2$.
%
%Furthermore, for any connected subword $t \leq w\!-\!1$.   Summing this over all connected subwords of an arbitrary word, results in $t \leq w\!-\!v\!-\!g\!-\!1$ and since we deal with a word with $v\!+\!g\geq 2$, clearly  $t < w\!-\!2$, so that a fortiori  $t\!-\!1<w\!-\!2$ hence $T\!-\!1<W\!-\!2$.
%
To find the upperbound of $T$, we note that the number of transpositions in a connected subword is strictly less than the width of that subword. This implies that  for the whole word  $t < w-v-g$, and since we deal with a word with at least one gap, $t<w-2 $ and hence  $T<W-2 $.
\hfill$\Box$
\item[Eq.  (\ref{PL3FFPR3})] can be treated in the same way, as it turns out that the contribution ${\rm PL_3 + FF + PR_3 }$ is equal to 
\be (-\tau)^g \Big[ \Delta_k(W\!-\!1,T) +  \Delta_k(W\!+\!1,T\!+\!1) \Big] \ee
Again we should verify if the arguments of $\Delta_k$ in both terms satisfy the restriction in (\ref{Delta}).   In this case the word need not have a gap, but it must have both an ascending and a descending sequence of monoids.   This implies that $1\leq t < w-1$, so that indeed the restrictions are satisfied in both terms.  \hfill$\Box$ 
\item[Eq.  (\ref{PL1FFPR1})] gets more involved, as ${\rm PL_1 + FF + PR_1 }$ is equal to \be (-\tau)^{g\!-\!1}\Big[\Delta_k(W\!-\!2,T\!-\!1)+\Delta_k(W\!-\!2,T)-\Delta_k(W\!-\!1,T)-\Delta_k(W\!+\!1,T\!+\!1)
\Big] \ee  
Each of the four terms vanish if $1 < T < W\!-\!2 $.
A word containing the environments coded in  (\ref{PL1FFPR1}) must have at least one gap of one vacancy, and a connected sequence with an ascending and descending part.   This implies that $v\!+\!g\geq 2$ and $t\geq 1$, so that $T = t \!+\! v \!+\! g > 2$, more than required.   The upper bound of $t$ is in these conditions $t\leq w\!-\!4$, so that the conditions are indeed fulfilled and  (\ref{PL1FFPR1}) is indeed satisfied.   \hfill$\Box$
\end{itemize}
This concludes the proof of the equations  (\ref{LRLRLR})-(\ref{PL1FFPR1}).

The solution of these equations can be expressed in four variables,  $\Lambda_1$, $\Lambda_2$, $\Lambda_3$, and $\Phi$ as follows
\be\def\arraycolsep{14pt}	
\begin{array}{llll}
\F\L_j =\Phi+\Lambda_j,&
\F\R_j=\Phi-\Lambda_j,&
\F\F=2\Phi,&
\L_j\R_k=\Lambda_j-\Lambda_k,\\
\P\L_j=-\Phi+\Lambda_j,&
\P\R_j=-\Phi-\Lambda_j,&
\P\P=-2\Phi&
\end{array}\label{solution}
\ee
It is straightforward to check that (\ref{solution}) satisfies the left-right anti-symmetry and solves the equations (\ref{LRLRLR}-\ref{PL1FFPR1}).
We remind the reader that we had six independent linear equations for ten independent contributions, so a four-parameter solution is complete.

Since every ascending and descending sequence has a beginning and an end, the number of P's is equal to the number of F's.   Likewise the entire word has a left- and a right-end, so that the number of L$_3$'s and of R$_3$'s are both equal to one.   Similarly each individual gap in the word, be it of one vacancy or longer, also has a right-- and a left--end, so that the number of   L$_1$'s and of R$_1$'s are equal as well as the number of  L$_2$'s and of R$_2$'s.  
Altogether this implies that $S_k(p)$ vanishes, and consequently that  $[Q_k,H] = 0$.

Before closing the section we note that the proof does not depend on the explicit form of the numbers $Z_k(w,t)$, (\ref{eq:Z-table}), but only on the triangle equation (\ref{triangle}).   We presented  (\ref{eq:Z-table}) in virtue of its closed form, its symmetry under $t\!+\!1 \leftrightarrow w-t $, and its maximal number of zero entries, which simplifies the explicit expressions for $Q_k$.
The most general solution of  (\ref{triangle}) has $k-1$ degrees of freedom for $Q_k$, and amounts to adding a numerical linear combination of $\tau^{j}\,Q_{k-j}$ to $Q_k$.
\section{Boundary conditions}
A useful variation on the periodic boundary conditions is the introduction of a twist $T$, by which 
\be  S^\al_{1+L} \equiv T_1 \,S^\al_{1}\, T_1^{-1}\qquad\text{ or equivalently}\qquad
S^\al_{1+L} \equiv R_{\al,\bt}\; S^\bt_{1} 
\ee
where the index on $T$ refers to the factor of $(\C^2)^{\otimes L}$ in which it acts.
If $R$ is a 3D rotation matrix then $T$ is its SU$_2$ representation.  This boundary condition on the Hamiltonian can be obtained from the transfer matrix by (\ref{HfromT}), where the transfer matrix $\TL$ is modified from equation (\ref{TM}) by the insertion of the twist matrix acting in the auxiliary space: 
\be  \TL^{\;\;T}(z) = {\rm Tr_a} \left(T_{\rm a} \prod_{j=1}^L R_{{\rm a},j}(z)\right)
\label{TMtwist} \ee
A priori the twist $T$ could be any invertible 2$\times$2 matrix, but without loss of generality for $H$ (though not for  $\TL^{\;\;T} (z)$) we choose $\det T=1$ from here on.
With this definition, the Hamiltonians (\ref{xxz-H}) and (\ref{TL-H})
generally differ by a term  $(q\!-\!q^{-1})(S_1^{\rm z} -T_1^{-1}\,S_1^{\rm z}\,T_1 )$.
This difference vanishes only when  $T$ is diagonal (commutes with $S^{\rm z}$).
Our $Q_j$ commute with the Hamiltonian $\sum_je_j$, (\ref{TL-H}).  This is equal to the XXZ Hamiltonian  (\ref{xxz-H}) with a twist $T$ only if $T$ is diagonal.
While the expressions for $Q_k$ are unaffected by the twist, the algebra is slightly altered, as $\tau' = {\rm Tr}\, T$ (after $T$ is normalized to $\det T=1$).
In Appendix \ref{twist} we argue that a non-diagonal $T$ does not act as a global twist, but as a localized scatterer.  Thus we conclude that for all {\it true twists}, and any value of $\tau$ and $\tau'$, our results are applicable.

An even more interesting variation on the boundary conditions is to open up the closed spin chain, and include specific boundary terms, such that the model remains integrable.  We are not aware of the existence of a boost operator~\cite{boost} to efficiently generate conserved quantities.  Differentiation of the (double-row) transfer matrix is still possible, but significantly more involved than in the (quasi-)periodic case.  An expected difference in the outcome is the existence of boundary terms in $Q_k$, in addition to the bulk terms which are the same as in the (quasi-)periodic case.  It is quite a challenge to see if our approach to find the complete structure works in this case.

\section{Conclusion}
We have given and proven a concise and efficient expression for the local conserved quantities of the closed XXZ chain, with or without a twist.  This may be instrumental in the investigation of equilibration of integrable models.
The conserved quantities are expressed in TL generators.  We have not attempted to express the result in the spin operators.  This can be done by plugging in the definition (equation~\ref{XXZmonoid}), but the resulting expressions may be rather large.  Some simplification is expected, in particular, the expressions in terms of spins must be even polynomials in terms of the spin operators $S_j^{\al}$, while the monoid itself is not.  Our present results are also applicable  to any other model, periodic or quasi-periodic, of which the Hamiltonian is a representation of (\ref{TL-H}), with the monoids satisfying (\ref{TL-rules}), without further restrictions.

We conjecture that there are no local conserved operators, independent of the series $Q_k$ presented here.  However, we do not have a proof for this claim.  It is clear that the transfer matrix is the generating function of the series $Q_k$.  We are not aware of a proof in the literature that there are no local conserved quantities (linearly) independent of those generated by the transfer matrix.  
%The question if there are non-local conserved quantities which are algebraically independent of $Q_k$ seems to be undecided at present.
The class of non-local conserved operators, see e.g.\cites{Halverson,Belletete}, seems more difficult.  The existence of  non-local conserved operators algebraically independent of $Q_k$ is not in doubt ($\rho$ is an example).  But their number and their properties are open questions.

Clearly the parallel result for the XYZ model \cite{Nozawa-Fukai} is more general, since the XXZ model is a special case of the XYZ model.
On the other hand our results apply to any other system which is similarly a representation of the (affine) TL algebra, of which there are many popular examples\cites{6vertex, BaxterEquivalence, TLPotts, WuPotts, roughening, Pasquier-ADE1, Pasquier-ADE2}.  For application of the results in calculations it is perhaps of greater interest that our result are extremely simple and compact and therefore easy to implement reliably, in contrast to \cite{Nozawa-Fukai}. 

\vspace{5mm}
\section*{Acknowledgements}
BN acknowledges many useful discussions with J.-S. Caux and V. Gritsev, and thanks  M.T. Batchelor and J. de Gier for the opportunity to present this material at Baxter2020 in Canberra.  A very helpful e-mail discussion with Filippo Colomo and Jean Michel Maillet on the effect of the twist, triggered us to investigate this in more detail.

\appendix
\section{Examples of $Q_k$}\label{examplesQk}
Here we give the first few examples of the conserved quantities, explicitly as a linear combination of words in the affine TL algebra.  The tables below show words in the TL algebra, and the coefficients of these words in the quantities  $Q_k$, equation (\ref{QkinCk}),  $A_k$ (or a multiple thereof), equation (\ref{Atilde2A}), and  $G_k$ defined recursively by $G_{k+1} = [B, G_k]$ with $G_1=H$, the Hamiltonian, and $B = \sum_j j\;e_j$ the boost operator, see \cite{boost}.  The words are given in two ways: the graphical form of the link pattern, the sequence of monoids in the product, represented as follows: $\sum_{j\in\Z_L} \prod_{n=1}^{m} e_{j + i_n}$ is written  as $[ i_m\;i_{m-1}\ldots i_2\;i_1 ]$.   The operator $A_k$ has fractional coefficients; and to avoid fractions in the table it is multiplied with the smallest common denominator (which is at most 4, for any $k$).

\def\pts{\hspace{1pt}}
\def\pa{\rct(0,0.5)(1,0.5)(1,0)}
\def\qa{\rct(0,-0.5)(-1,-0.5)(-1,0)}
\def\pb{\rct(0,0.6)(2,1.0)(2,1.6)}
\def\qb{\rct(0,0.6)(-2,1.0)(-2,1.6)}
\def\pc{\rct(0,1.1)(3,1.1)(3,0)}
\def\qc{\rct(0,-1.1)(-3,-1.1)(-3,0)}
\def\pd{\rct(0,1.4)(4,0.2)(4,1.6)}
\def\qd{\rct(0,1.4)(-4,0.2)(-4,1.6)}
% k,w,t,v,g:  2
\begin{table}[hb]\caption
%  \newpage\noindent Table 5:
{ List of words and their coefficients in the quantities $Q_k$, $A_k$ and $G_k$ for $k\in\{2,3,4\}$. The words are represented by a link pattern, and by a sequence of monoid indices.}
\begin{center}
\renewcommand{\arraystretch}{1.2}
\psset{unit=3mm}
\begin{tabular}{|c|c|r|r|r|}\hline \hline
link p. & indices & $Q_2$ & $A_2$ & $G_2$ \\\hline  
% k,w,t,v,g:  2  2  0  0  0      19 %
\pspicture(0,0)(2,2)\pscustom{   \mt(0,0)\pb  \mt(1,0)\pa  \mt(1,1.6)\qa  }\endpspicture & [\pts0\pts1\pts] &%
$2$ & $1$ & $1$ \\
% k,w,t,v,g:  2  2  1  0  0      13 %
\pspicture(0,0)(2,1.6)\pscustom{   \mt(0,0)\pa  \mt(2,0)\qb  \mt(2,1.6)\qa  }\endpspicture & [\pts1\pts0\pts] &%
$-2$ & $-1$ & $-1$ \\
% k,w,t,v,g:  3    
 \hline \multicolumn{5}{c}{} \\
 \multicolumn{5}{c}{} \\ \hline \hline
link p. & indices & $Q_3$ & $2A_3$ & $G_3$ \\\hline  
% k,w,t,v,g:  3  2  0  0  0      75 %
\pspicture(0,0)(2,2)\pscustom{  %
\mt(0,0)\pb %
\mt(1,0)\pa %
\mt(1,1.6)\qa %
}\endpspicture &%
[\pts0\pts1\pts] &%
$\tau$ &%
$\tau$ &%
$\tau$ \\
% k,w,t,v,g:  3  2  1  0  0      45 %
\pspicture(0,0)(2,1.6)\pscustom{  %
\mt(0,0)\pa %
\mt(2,0)\qb %
\mt(2,1.6)\qa %
}\endpspicture &%
[\pts1\pts0\pts] &%
$\tau$ &%
$\tau$ &%
$\tau$ \\
% k,w,t,v,g:  3  3  0  0  0      71 %
\pspicture(0,0)(3,1.6)\pscustom{  %
\mt(0,0)\pb %
\mt(1,0)\pb %
\mt(2,0)\pa %
\mt(1,1.6)\qa %
}\endpspicture &%
[\pts0\pts1\pts2\pts] &%
$2$ &%
$2$ &%
$2$ \\
% k,w,t,v,g:  3  3  2  0  0      29 %
\pspicture(0,0)(3,1.6)\pscustom{  %
\mt(0,0)\pa %
\mt(2,0)\qb %
\mt(3,0)\qb %
\mt(3,1.6)\qa %
}\endpspicture &%
[\pts2\pts1\pts0\pts] &%
$2$ &%
$2$ &%
$2$ \\
% k,w,t,v,g:  3  3  1  0  0      53 %
\pspicture(0,0)(3,1.6)\pscustom{  %
\mt(0,0)\pa %
\mt(2,0)\pa %
\mt(3,1.6)\qc %
\mt(2,1.6)\qa %
}\endpspicture &%
[\pts1\pts0\pts2\pts] &%
$-2$ &%
$-2$ &%
$-2$ \\
% k,w,t,v,g:  3  3  1  0  0      83 %
\pspicture(0,0)(3,1.6)\pscustom{  %
\mt(0,0)\pc %
\mt(1,0)\pa %
\mt(3,1.6)\qa %
\mt(1,1.6)\qa %
}\endpspicture &%
[\pts0\pts2\pts1\pts] &%
$-2$ &%
$-2$ &%
$-2$ \\
% k,w,t,v,g:  4    
   \hline\end{tabular} \qquad
\renewcommand{\arraystretch}{1.05}
 \begin{tabular}{|c|c|r|r|r|}\hline \hline
link p. & indices & $Q_4$ \hspace*{1mm}& $A_4$\hspace*{6mm}  & $G_4$\hspace*{5mm} \\\hline  
% k,w,t,v,g:  4  2  0  0  0     283 %
\pspicture(0,0)(2,2)\pscustom{  %
\mt(0,0)\pb %
\mt(1,0)\pa %
\mt(1,1.6)\qa %
}\endpspicture &%
[\pts0\pts1\pts] &%
$-2$ &%
$-1\!-\!\tau^2$ &%
$2\!+\!\tau^2$ \\
% k,w,t,v,g:  4  2  1  0  0     157 %
\pspicture(0,0)(2,1.6)\pscustom{  %
\mt(0,0)\pa %
\mt(2,0)\qb %
\mt(2,1.6)\qa %
}\endpspicture &%
[\pts1\pts0\pts] &%
$2$ &%
$1\!+\!\tau^2$ &%
$-2\!-\!\tau^2$ \\
% k,w,t,v,g:  4  3  0  0  0     279 %
\pspicture(0,0)(3,1.6)\pscustom{  %
\mt(0,0)\pb %
\mt(1,0)\pb %
\mt(2,0)\pa %
\mt(1,1.6)\qa %
}\endpspicture &%
[\pts0\pts1\pts2\pts] &%
$2\tau$ &%
$\tau$ &%
$6\tau$ \\
% k,w,t,v,g:  4  3  2  0  0      93 %
\pspicture(0,0)(3,1.6)\pscustom{  %
\mt(0,0)\pa %
\mt(2,0)\qb %
\mt(3,0)\qb %
\mt(3,1.6)\qa %
}\endpspicture &%
[\pts2\pts1\pts0\pts] &%
$-2\tau$ &%
$-\tau$ &%
$-6\tau$ \\
% k,w,t,v,g:  4  4  0  0  0     271 %
\pspicture(0,0)(4,1.6)\pscustom{  %
\mt(0,0)\pb %
\mt(1,0)\pb %
\mt(2,0)\pb %
\mt(3,0)\pa %
\mt(1,1.6)\qa %
}\endpspicture &%
[\pts0\pts1\pts2\pts3\pts] &%
$2$ &%
$1$ &%
$6$ \\
% k,w,t,v,g:  4  4  2  0  0     109 %
\pspicture(0,0)(4,1.6)\pscustom{  %
\mt(0,0)\pa %
\mt(2,0)\qb %
\mt(3,0)\pa %
\mt(4,1.6)\qc %
\mt(3,1.6)\qa %
}\endpspicture &%
[\pts2\pts1\pts0\pts3\pts] &%
$2$ &%
$1$ &%
$6$ \\
% k,w,t,v,g:  4  4  2  0  0     181 %
\pspicture(0,0)(4,1.6)\pscustom{  %
\mt(0,0)\pa %
\mt(2,0)\pa %
\mt(4,0)\qd %
\mt(4,1.6)\qa %
\mt(2,1.6)\qa %
}\endpspicture &%
[\pts1\pts0\pts3\pts2\pts] &%
$2$ &%
$1$ &%
$6$ \\
\hline\end{tabular}
 \end{center} \end{table}%\\[6mm]
 \begin{table}[ht]\caption
%Table 6:
{List of words and their coefficients in the quantities $Q_5$, $A_5$ and $G_5$.
To save space pairs of words with the same coefficient are placed in the same row.}
\begin{center}
\renewcommand{\arraystretch}{1.3}\psset{unit=3mm}
\begin{tabular}{|rl|rl||r|r|r|}\hline \hline link pattern & indices & link pattern & indices &$Q_5$\hspace*{5mm}&$2 A_5$\hspace*{5mm} &$G_5$\hspace*{7mm} \\ \hline
\pspicture(0,0)(2,2)\pscustom{  \mt(0,0)\pb 	\mt(1,0)\pa 	\mt(1,1.6)\qa 	}\endpspicture &	[\pts0\pts1\pts] &
\pspicture(0,0)(2,1.6)\pscustom{  \mt(0,0)\pa 	\mt(2,0)\qb 	\mt(2,1.6)\qa 	}\endpspicture &	[\pts1\pts0\pts] &	$-\tau$ &	$-\tau\!-\!2\tau^3$ &	$8\tau\!+\!\tau^3$ \\
&&\pspicture(0,0)(3,1.6)\pscustom{  \mt(0,0)\pa 	\mt(2,0)\pa 	\mt(3,1.6)\qa 	\mt(1,1.6)\qa 	}\endpspicture &	[\pts0\pts2\pts] & 	$-2\tau$ &	$-2\tau$ &	$-24\tau$ \\
\pspicture(0,0)(3,1.6)\pscustom{  \mt(0,0)\pb 	\mt(1,0)\pb 	\mt(2,0)\pa 	\mt(1,1.6)\qa 	}\endpspicture &	[\pts0\pts1\pts2\pts] &	
\pspicture(0,0)(3,1.6)\pscustom{  \mt(0,0)\pa 	\mt(2,0)\qb 	\mt(3,0)\qb 	\mt(3,1.6)\qa 	}\endpspicture &	[\pts2\pts1\pts0\pts] &	$-2\!+\!\tau^2$ &	$-2\!-\!3\tau^2$ &	$16\!+\!14\tau^2$ \\
\pspicture(0,0)(3,1.6)\pscustom{  \mt(0,0)\pa 	\mt(2,0)\pa 	\mt(3,1.6)\qc 	\mt(2,1.6)\qa 	}\endpspicture &	[\pts1\pts0\pts2\pts] &	
\pspicture(0,0)(3,1.6)\pscustom{  \mt(0,0)\pc 	\mt(1,0)\pa 	\mt(3,1.6)\qa 	\mt(1,1.6)\qa 	}\endpspicture &	[\pts0\pts2\pts1\pts] &	$2$ &	$2\!+\!4\tau^2$ &	$-16\!-\!2\tau^2$ \\
\pspicture(0,0)(4,1.6)\pscustom{  \mt(0,0)\pb 	\mt(1,0)\pb 	\mt(2,0)\pb 	\mt(3,0)\pa 	\mt(1,1.6)\qa 	}\endpspicture &	[\pts0\pts1\pts2\pts3\pts] &
\pspicture(0,0)(4,1.6)\pscustom{  \mt(0,0)\pa 	\mt(2,0)\qb 	\mt(3,0)\qb 	\mt(4,0)\qb 	\mt(4,1.6)\qa 	}\endpspicture &	[\pts3\pts2\pts1\pts0\pts] &	$3\tau$ &	$3\tau$ &	$36\tau$ \\
\pspicture(0,0)(4,1.6)\pscustom{  \mt(0,0)\pa 	\mt(2,0)\pb 	\mt(3,0)\pa 	\mt(3,1.6)\qc 	\mt(2,1.6)\qa 	}\endpspicture &	[\pts1\pts0\pts2\pts3\pts] &
\pspicture(0,0)(4,1.6)\pscustom{  \mt(0,0)\pa 	\mt(2,0)\qb 	\mt(3,0)\pa 	\mt(4,1.6)\qc 	\mt(3,1.6)\qa 	}\endpspicture &	[\pts2\pts1\pts0\pts3\pts] &	$-\tau$ &	$-\tau$ &	$-12\tau$ \\
\hline\end{tabular}\end{center}\end{table}
\clearpage
\begin{table}[h!]\caption
%\noindent Table 7:
{List of words and their coefficients in the quantities $Q_6$, $A_6$ and $G_6$. Two or four words with the same coefficients are listed on the same line}
%\\[2mm]\samepage
\renewcommand{\arraystretch}{1.2}
\begin{center}
\psset{unit=3mm}
\noindent	 \begin{tabular}{|rl|rl||r|r|r|}\hline
link p. & indices & link p. & indices & $Q_6$\hspace{3mm} & $ A_6$\hspace{1cm} & $G_6$\hspace*{1cm}\\ \hline
&&\pspicture(0,0)(1.7,2)\pscustom{ \mt(0,0.0)\pb 	\mt(1,0.0)\pa 	\mt(1,1.6)\qa 	}\endpspicture &	[\,0\,1\,]  &	$2$ &	$1\!+\!3\tau^2\!+\!3\tau^4$ &	$16\!+\!22\tau^2\!+\!\tau^4$ \\
&&\pspicture(0,0)(1.7,1.6)\pscustom{ \mt(0,0.0)\pa 	\mt(2,0.0)\qb 	\mt(2,1.6)\qa 	}\endpspicture &	[\,1\,0\,]  &	$-2$ &	$-1\!-\!3\tau^2\!-\!3\tau^4$ &	$-16\!-\!22\tau^2\!-\!\tau^4$ \\
&&\pspicture(0,0)(2.7,1.6)\pscustom{ \mt(0,0.0)\pb 	\mt(1,0.0)\pb 	\mt(2,0.0)\pa 	\mt(1,1.6)\qa 	}\endpspicture &	[\,0\,1\,2\,] &	$-2\tau$ &	$-\tau\!-\!3\tau^3$ &	$120\tau\!+\!30\tau^3$ \\
&&\pspicture(0,0)(2.7,1.6)\pscustom{ \mt(0,0.0)\pa 	\mt(2,0.0)\qb 	\mt(3,0.0)\qb 	\mt(3,1.6)\qa 	}\endpspicture &	[\,2\,1\,0\,] &	$2\tau$ &	$\tau\!+\!3\tau^3$ &	$-120\tau\!-\!30\tau^3$ \\
&&\pspicture(0,0)(3.7,1.6)\pscustom{ \mt(0,0.0)\pb 	\mt(1,0.0)\pb 	\mt(2,0.0)\pb 	\mt(3,0.0)\pa 	\mt(1,1.6)\qa 	}\endpspicture &	[\,0\,1\,2\,3\,] &	$-2\!+\!2\tau^2$ &	$-1\!-\!2\tau^2$ &	$120\!+\!150\tau^2$ \\
&&\pspicture(0,0)(3.7,1.6)\pscustom{ \mt(0,0.0)\pa 	\mt(2,0.0)\qb 	\mt(3,0.0)\qb 	\mt(4,0.0)\qb 	\mt(4,1.6)\qa 	}\endpspicture &	[\,3\,2\,1\,0\,] &	$2\!-\!2\tau^2$ &	$1\!+\!2\tau^2$ &	$-120\!-\!150\tau^2$ \\
	\pspicture(0,0)(3.7,1.6)\pscustom{ \mt(0,0.0)\pb 	\mt(1,0.0)\pa 	\mt(3,0.0)\pa 	\mt(4,1.6)\qa 	\mt(1,1.6)\qa 	}\endpspicture &	[\,0\,1\,3\,] &	
	\pspicture(0,0)(3.7,1.6)\pscustom{ \mt(0,0.0)\pa 	\mt(2,0.0)\pb 	\mt(3,0.0)\pa 	\mt(3,1.6)\qa 	\mt(1,1.6)\qa 	}\endpspicture &	[\,0\,2\,3\,] &	$-2\tau$ &	$-\tau$ &	$-120\tau$ \\
\hline\end{tabular} \\ \psset{unit=2mm}\setlength{\tabcolsep}{5.6pt}
	\begin{tabular}{|cccccccc||r|r|r|} \multicolumn{2}{|c}{continued:} \\ \hline 
	\multicolumn{8}{|l||}{pairs of link patterns and index sequences } &\!$Q_6$\!&\!$A_6$\!& $G_6$\\ \hline
	\pspicture(0,0)(5,1.6)\pscustom{ \mt(0,0.0)\pb 	\mt(1,0.0)\pb 	\mt(2,0.0)\pb 	\mt(3,0.0)\pb 	\mt(4,0.0)\pb 	\mt(5,0.0)\pa 	\mt(1,1.6)\qa 		}\endpspicture &	[012345] &
	\pspicture(-.2,0)(5,1.6)\pscustom{ \mt(0,0.0)\pa 	\mt(2,0.0)\qb 	\mt(3,0.0)\pb 	\mt(4,0.0)\pb 	\mt(5,0.0)\pa 	\mt(4,1.6)\qc 	\mt(3,1.6)\qa 		}\endpspicture &	[210345] &
	\pspicture(-.2,0)(5,1.6)\pscustom{ \mt(0,0.0)\pa\mt(2,0.0)\pa\mt(4,0.0)\pb\mt(5,0.0)\pa\mt(5,1.6)\rct(0,-1.1)(-5,-1.1)(-5,0.0)\mt(4,1.6)\qa\mt(2,1.6)\qa	}\endpspicture &	[103245] &
	\pspicture(-.2,0)(5,1.6)\pscustom{ \mt(0,0.0)\pa 	\mt(2,0.0)\pb 	\mt(3,0.0)\pc 	\mt(4,0.0)\pa 	\mt(6,1.6)\qa 	\mt(3,1.6)\qc 	\mt(2,1.6)\qa 		}\endpspicture &	[102354] &	2 &	1 &	120 \\
	\pspicture(0,0)(5,1.6)\pscustom{ \mt(0,0.0)\pa 	\mt(2,0.0)\pd 	\mt(3,0.0)\pa 	\mt(5,0.0)\pa 	\mt(5,1.6)\qa 	\mt(3,1.6)\qc 	\mt(2,1.6)\qa 		}\endpspicture &	[102435] &
	\pspicture(-.2,0)(5,1.6)\pscustom{ \mt(0,0.0)\pb 	\mt(1,0.0)\pb 	\mt(2,0.0)\pc 	\mt(3,0.0)\pa 	\mt(6,0.0)\qb 	\mt(6,1.6)\qa 	\mt(1,1.6)\qa 		}\endpspicture &	[012543] &
	\pspicture(-.2,0)(5,1.6)\pscustom{ \mt(0,0.0)\pb 	\mt(1,0.0)\pc 	\mt(2,0.0)\pa 	\mt(5,0.0)\pa 	\mt(6,1.6)\qc 	\mt(5,1.6)\qa 	\mt(1,1.6)\qa 		}\endpspicture &	[014325] &
	\pspicture(-.2,0)(5,1.6)\pscustom{ \mt(0,0.0)\pb\mt(1,0.0)\rct(0,1.1)(5,1.1)(5,0.0)\mt(2,0.0)\pa\mt(4,0.0)\pa\mt(6,1.6)\qa\mt(4,1.6)\qa \mt(1,1.6)\qa	}\endpspicture &	[013254] &	2 &	1 &	120 \\
\hline\end{tabular} \end{center}
 \end{table}
\clearpage
% \begin{table}[h!]
\noindent Table 8:
%\caption
{List of words and their coefficients in the quantities $Q_7$, $A_7$ and $G_7$.}
\\[1mm]
  \setlength{\tabcolsep}{5pt}
\psset{unit=2mm}\def\pts{\hspace{1pt}}
\begin{tabular}{|rl|rl||r|r|r|}\hline
link p.  &  indices  & link p.  &  indices  & $Q_7$\hspace*{5mm}& $4 A_7$\hspace*{11mm}&  $G_7$\hspace*{13mm} \\\hline
\pspicture(0,0)(2,2)\pscustom{ \mt(0,0.0)\pb 	\mt(1,0.0)\pa 	\mt(1,1.6)\qa 	}\endpspicture &	[\pts0\pts1\pts] &
\pspicture(0,0)(2,1.6)\pscustom{ \mt(0,0.0)\pa 	\mt(2,0.0)\qb 	\mt(2,1.6)\qa 	}\endpspicture &	[\pts1\pts0\pts] &	$\tau$ &	$2\tau\!+\!9\tau^3\!+\!17\tau^5$ &	$136\tau\!+\!52\tau^3\!+\!\tau^5$ \\
&&\pspicture(0,0)(3,1.6)\pscustom{ \mt(0,0.0)\pa 	\mt(2,0.0)\pa 	\mt(3,1.6)\qa 	\mt(1,1.6)\qa 	}\endpspicture &	[\pts0\pts2\pts] &	$2\tau$ &	$4\tau\!+\!18\tau^3$ &	$\!-\!960\tau\!-\!120\tau^3$ \\
&&\pspicture(0,0)(4,1.6)\pscustom{ \mt(0,0.0)\pa 	\mt(2,0)\rlt(0,1.6) 	\mt(3,0.0)\pa 	\mt(4,1.6)\qa 	\mt(1,1.6)\qa 	}\endpspicture &	[\pts0\pts3\pts] &	$2\tau$ &	$4\tau$ &	$720\tau$ \\
\pspicture(0,0)(3,1.6)\pscustom{ \mt(0,0.0)\pb 	\mt(1,0.0)\pb 	\mt(2,0.0)\pa 	\mt(1,1.6)\qa 	}\endpspicture &	[\pts0\pts1\pts2\pts] &
\pspicture(0,0)(3,1.6)\pscustom{ \mt(0,0.0)\pa 	\mt(2,0.0)\qb 	\mt(3,0.0)\qb 	\mt(3,1.6)\qa 	}\endpspicture &	[\pts2\pts1\pts0\pts] &	$2\!-\!\tau^2$ &	$4\!+\!16\tau^2\!+\!25\tau^4$ &	$272\!+\!584\tau^2\!+\!62\tau^4$ \\
\pspicture(0,0)(3,1.6)\pscustom{ \mt(0,0.0)\pa 	\mt(2,0.0)\pa 	\mt(3,1.6)\qc 	\mt(2,1.6)\qa 	}\endpspicture &	[\pts1\pts0\pts2\pts] &
\pspicture(0,0)(3,1.6)\pscustom{ \mt(0,0.0)\pc 	\mt(1,0.0)\pa 	\mt(3,1.6)\qa 	\mt(1,1.6)\qa 	}\endpspicture &	[\pts0\pts2\pts1\pts] &	$\!-\!2$ &	$\!-\!4\!-\!18\tau^2\!-\!34\tau^4$ &	$\!-\!272\!-\!104\tau^2\!-\!2\tau^4$ \\
\pspicture(0,0)(4,1.6)\pscustom{ \mt(0,0.0)\pb 	\mt(1,0.0)\pa 	\mt(3,0.0)\pa 	\mt(4,1.6)\qa 	\mt(1,1.6)\qa 	}\endpspicture &	[\pts0\pts1\pts3\pts] &
\pspicture(0,0)(4,1.6)\pscustom{ \mt(0,0.0)\pa 	\mt(2,0.0)\pb 	\mt(3,0.0)\pa 	\mt(3,1.6)\qa 	\mt(1,1.6)\qa 	}\endpspicture &	[\pts0\pts2\pts3\pts] &	$\!-\!\tau^2$ &	$\!-\!2\tau^2$ &	$\!-\!360\tau^2$ \\
\hline\end{tabular} \\[5mm]
The following words have the coefficients $\tau$, $\tau/2$, and $360\tau$ in $Q_7$, $A_7$ and $G_7$ respectively:\\[1mm]\samepage\setlength{\tabcolsep}{6pt}
\begin{tabular}{cccccccc}
\pspicture(0,0)(6,1.6)\pscustom{ \mt(0,0.0)\pa 	\mt(2,0.0)\qb 	\mt(3,0.0)\pb 	\mt(4,0.0)\pb 	\mt(5,0.0)\pa 	\mt(4,1.6)\qc 	\mt(3,1.6)\qa 	}\endpspicture &	[\pts2\pts1\pts0\pts3\pts4\pts5\pts] &
\pspicture(0,0)(6,1.6)\pscustom{ \mt(0,0.0)\pa 	\mt(2,0.0)\pb 	\mt(3,0.0)\pc 	\mt(4,0.0)\pa 	\mt(6,1.6)\qa 	\mt(3,1.6)\qc 	\mt(2,1.6)\qa 	}\endpspicture &	[\pts1\pts0\pts2\pts3\pts5\pts4\pts] &
\pspicture(0,0)(6,1.6)\pscustom{ \mt(0,0.0)\pa 	\mt(2,0.0)\pd 	\mt(3,0.0)\pa 	\mt(5,0.0)\pa 	\mt(5,1.6)\qa 	\mt(3,1.6)\qc 	\mt(2,1.6)\qa 	}\endpspicture &	[\pts1\pts0\pts2\pts4\pts3\pts5\pts] &
\pspicture(0,0)(6,1.6)\pscustom{ \mt(0,0.0)\pb 	\mt(1,0.0)\pb 	\mt(2,0.0)\pc 	\mt(3,0.0)\pa 	\mt(6,0.0)\qb 	\mt(6,1.6)\qa 	\mt(1,1.6)\qa 	}\endpspicture &	[\pts0\pts1\pts2\pts5\pts4\pts3\pts] \\
\pspicture(0,0)(6,1.6)\pscustom{ \mt(0,0.0)\pb 	\mt(1,0.0)\pc 	\mt(2,0.0)\pa 	\mt(5,0.0)\pa 	\mt(6,1.6)\qc 	\mt(5,1.6)\qa 	\mt(1,1.6)\qa 	}\endpspicture &	[\pts0\pts1\pts4\pts3\pts2\pts5\pts] &
\pspicture(0,0)(6,1.6)\pscustom{ \mt(0,0.0)\pc 	\mt(1,0.0)\pa 	\mt(4,0.0)\pb 	\mt(5,0.0)\pa 	\mt(5,1.6)\qc 	\mt(4,1.6)\qa 	\mt(1,1.6)\qa 	}\endpspicture &	[\pts0\pts3\pts2\pts1\pts4\pts5\pts] &
\pspicture(0,0)(6,1.6)\pscustom{ \mt(0,0.0)\pd 	\mt(1,0.0)\pa 	\mt(3,0.0)\pc 	\mt(4,0.0)\pa 	\mt(6,1.6)\qa 	\mt(3,1.6)\qa 	\mt(1,1.6)\qa 	}\endpspicture &	[\pts0\pts2\pts1\pts3\pts5\pts4\pts] &
\pspicture(0,0)(6,1.6)\pscustom{ \mt(0,0.0)\pa 	\mt(2,0.0)\qb 	\mt(3,0.0)\qb 	\mt(4,0.0)\pb 	\mt(5,0.0)\pa 	\mt(5,1.6)\qc 	\mt(4,1.6)\qa 	}\endpspicture &	[\pts3\pts2\pts1\pts0\pts4\pts5\pts] \\
\end{tabular} \\[3mm]
\newpage\noindent
 The following words have the coefficients 2, 1, and 720 in $Q_7$, $A_7$ and $G_7$ respectively:
\def\pts{\hspace{.5pt}}
\samepage\\[1mm]
\samepage\begin{tabular}{cccccccc}
\pspicture(0,0)(6.5,1.6)\pscustom{ \mt(0,0.0)\pa 	\mt(2,0.0)\pa 	\mt(4,0.0)\qd 	\mt(5,0.0)\pa 	\mt(7,0.0)\qd 	\mt(7,1.6)\qa 	\mt(5,1.6)\qa 	\mt(2,1.6)\qa 	}\endpspicture &	[\pts1\pts0\pts4\pts3\pts2\pts6\pts5\pts] &
\pspicture(-.5,0)(6.5,1.6)\pscustom{ \mt(0,0.0)\pa 	\mt(2,0.0)\pa 	\mt(4,0.0)\qd 	\mt(5,0.0)\qb 	\mt(6,0.0)\pa 	\mt(7,1.6)\qc 	\mt(6,1.6)\qa 	\mt(2,1.6)\qa 	}\endpspicture &	[\pts1\pts0\pts5\pts4\pts3\pts2\pts6\pts] &
\pspicture(-.5,0)(6.5,1.6)\pscustom{ \mt(0,0.0)\pa 	\mt(2,0.0)\pb 	\mt(3,0.0)\pb 	\mt(4,0.0)\pc 	\mt(5,0.0)\pa 	\mt(7,1.6)\qa 	\mt(3,1.6)\qc 	\mt(2,1.6)\qa 	}\endpspicture &	[\pts1\pts0\pts2\pts3\pts4\pts6\pts5\pts] &
\pspicture(-.5,0)(6.5,1.6)\pscustom{ \mt(0,0.0)\pa 	\mt(2,0.0)\pb 	\mt(3,0.0)\pd 	\mt(4,0.0)\pa 	\mt(6,0.0)\pa 	\mt(6,1.6)\qa 	\mt(3,1.6)\qc 	\mt(2,1.6)\qa 	}\endpspicture &	[\pts1\pts0\pts2\pts3\pts5\pts4\pts6\pts] \\
\pspicture(0,0)(6.5,1.6)\pscustom{ \mt(0,0.0)\pa 	\mt(2,0.0)\pc 	\mt(3,0.0)\pa 	\mt(6,0.0)\qb 	\mt(7,0.0)\qb 	\mt(7,1.6)\qa 	\mt(3,1.6)\qc 	\mt(2,1.6)\qa 	}\endpspicture &	[\pts1\pts0\pts2\pts6\pts5\pts4\pts3\pts] &
\pspicture(-.5,0)(6.5,1.6)\pscustom{ \mt(0,0.0)\pa 	\mt(2,0.0)\pd 	\mt(3,0.0)\pa 	\mt(5,0.0)\pb 	\mt(6,0.0)\pa 	\mt(5,1.6)\qa 	\mt(3,1.6)\qc 	\mt(2,1.6)\qa 	}\endpspicture &	[\pts1\pts0\pts2\pts4\pts3\pts5\pts6\pts] &
\pspicture(-.5,0)(6.5,1.6)\pscustom{ \mt(0,0.0)\pa 	\mt(2,0.0)\qb 	\mt(3,0.0)\pa 	\mt(5,0.0)\qd 	\mt(6,0.0)\pa 	\mt(7,1.6)\qc 	\mt(6,1.6)\qa 	\mt(3,1.6)\qa 	}\endpspicture &	[\pts2\pts1\pts0\pts5\pts4\pts3\pts6\pts] &
\pspicture(-.5,0)(6.5,1.6)\pscustom{ \mt(0,0.0)\pa 	\mt(2,0.0)\qb 	\mt(3,0.0)\pb 	\mt(4,0.0)\pb 	\mt(5,0.0)\pb 	\mt(6,0.0)\pa 	\mt(4,1.6)\qc 	\mt(3,1.6)\qa 	}\endpspicture &	[\pts2\pts1\pts0\pts3\pts4\pts5\pts6\pts] \\
\end{tabular}
 \\[5mm]
The following words have the coefficients -2, -1, and -720 in $Q_7$, $A_7$ and $G_7$ respectively:\\[1mm]
\samepage \begin{tabular}{cccccccc}
\pspicture(0,0)(6.5,1.6)\pscustom{ \mt(0,0.0)\pa 	\mt(2,0.0)\pa 	\mt(4,0.0)\qd 	\mt(5,0.0)\pb 	\mt(6,0.0)\pa 	\mt(6,1.6)\qc 	\mt(5,1.6)\qa 	\mt(2,1.6)\qa 	}\endpspicture &	[\pts1\pts0\pts4\pts3\pts2\pts5\pts6\pts] &
\pspicture(-.5,0)(6.5,1.6)\pscustom{ \mt(0,0.0)\pa 	\mt(2,0.0)\pa 	\mt(4,0.0)\qd 	\mt(5,0.0)\qb 	\mt(6,0.0)\qb 	\mt(7,0.0)\qb 	\mt(7,1.6)\qa 	\mt(2,1.6)\qa 	}\endpspicture &	[\pts1\pts0\pts6\pts5\pts4\pts3\pts2\pts] &
\pspicture(-.5,0)(6.5,1.6)\pscustom{ \mt(0,0.0)\pa 	\mt(2,0.0)\pb 	\mt(3,0.0)\pb 	\mt(4,0.0)\pb 	\mt(5,0.0)\pb 	\mt(6,0.0)\pa 	\mt(3,1.6)\qc 	\mt(2,1.6)\qa 	}\endpspicture &	[\pts1\pts0\pts2\pts3\pts4\pts5\pts6\pts] &
\pspicture(-.5,0)(6.5,1.6)\pscustom{ \mt(0,0.0)\pa 	\mt(2,0.0)\pb 	\mt(3,0.0)\pc 	\mt(4,0.0)\pa 	\mt(7,0.0)\qb 	\mt(7,1.6)\qa 	\mt(3,1.6)\qc 	\mt(2,1.6)\qa 	}\endpspicture &	[\pts1\pts0\pts2\pts3\pts6\pts5\pts4\pts] \\
\pspicture(0,0)(6.5,1.6)\pscustom{ \mt(0,0.0)\pa 	\mt(2,0.0)\pc 	\mt(3,0.0)\pa 	\mt(6,0.0)\pa 	\mt(7,1.6)\qc 	\mt(6,1.6)\qa 	\mt(3,1.6)\qc 	\mt(2,1.6)\qa 	}\endpspicture &	[\pts1\pts0\pts2\pts5\pts4\pts3\pts6\pts] &
\pspicture(-.5,0)(6.5,1.6)\pscustom{ \mt(0,0.0)\pa 	\mt(2,0.0)\qb 	\mt(3,0.0)\pa 	\mt(5,0.0)\qd 	\mt(6,0.0)\qb 	\mt(7,0.0)\qb 	\mt(7,1.6)\qa 	\mt(3,1.6)\qa 	}\endpspicture &	[\pts2\pts1\pts0\pts6\pts5\pts4\pts3\pts] &
\pspicture(-.5,0)(6.5,1.6)\pscustom{ \mt(0,0.0)\pa 	\mt(2,0.0)\qb 	\mt(3,0.0)\pb 	\mt(4,0.0)\pc 	\mt(5,0.0)\pa 	\mt(7,1.6)\qa 	\mt(4,1.6)\qc 	\mt(3,1.6)\qa 	}\endpspicture &	[\pts2\pts1\pts0\pts3\pts4\pts6\pts5\pts] &
\pspicture(-.5,0)(6.5,1.6)\pscustom{ \mt(0,0.0)\pa 	\mt(2,0.0)\qb 	\mt(3,0.0)\pd 	\mt(4,0.0)\pa 	\mt(6,0.0)\pa 	\mt(6,1.6)\qa 	\mt(4,1.6)\qc 	\mt(3,1.6)\qa 	}\endpspicture &	[\pts2\pts1\pts0\pts3\pts5\pts4\pts6\pts] \\
\end{tabular}%\end{table}

\section{A true twist}\label{twist}
Consider a closed quantum chain, with site operators $S_j$, with $j\in\Z_L$, acting in the $j$-th factor of ${\cal H}_L:={\cal V}^{\otimes L}$.    Introduce a general twist to the periodic boundary conditions, described by a matrix $T$.   One way to implement such twist is to write the Hamiltonian as \be  H = \sum_{j=1}^L h(S_j,S_{j+1})\,,\ee while identifying the operators $S_{L+1} \equiv T_1\,S_1\, T_1^{-1}$.   The local interaction $h$ is some simple expression in its arguments.

One can also ignore $S_{L+1}$, and write the interaction of $S_L$ with $S_1$ explicitly: 
\be   H_1 =  h\left(S_L, T_1 S_1  T_1^{-1}\right) + \sum_{j=2}^L h(S_{j-1}, S_j)\,,  \label{H1}\ee 
a formulation which is a special case of modeling an impurity locally affecting the interaction.   This raises the question under what conditions such a variant interaction in one position can be interpreted as a twist, rather than an impurity.
We find it natural to reserve the word twist for the case 
that a local observable is unchanged by changing the locus of the twist as long as it does not pass through the position of the local observable.  If this is not the case, one can effectively measure the distance to the impurity.  We are aware that we now use the word twist in a more limited sense than in the literature (e.g. \cite{twisted-XXZ}), where the twist simply means any coordinate transformation before identifying the $L\!+\!1$-st and 1-st factor space.  Therefore we use the phrase {\it true twist}, to refer to a twist in this more limited sense.
Note that the question whether the model is integrable is a different one: an impurity does not automatically violate integrablity.

Consider a non-degenerate eigenstate $\psi$ of  Hamiltonian $H_1$ (\ref{H1}).  It is a vector in ${\cal H}_L$, so one may endow it with $L$ indices.   A translation operator $A$ which cycles the factor spaces around can be written as $\left(A\, \psi\right)_{n_1,n_2,\ldots,n_L}=\psi_{n_2,n_3,\ldots,n_{L},n_1}  $.   But indices will be suppressed where possible.
\be  H_1 \psi = E \psi \ee
Then what is the corresponding eigenstate of a modified Hamiltonian, $H_2$, in which the position of the exceptional interaction is shifted?
\be   H_2 =A H_1 A^{-1} =  h(S_L,S_1) + h\left(S_1, T_2 S_2  T_2^{-1}\right) + \sum_{j=3}^L h(S_{j-1}, S_j)\ee
Clearly
\be  H_2 (A \psi) = A H_1 A^{-1} A\psi = A H_1 \psi = E \left(A \psi\right)\,, \ee which shows that $A\,\psi$ is the eigenstate of $H_2$ that corresponds with $\psi$.   
The question now is if $\psi$ and $A\,\psi$ differ only locally.   It is suggestive to try $A\psi \propto T_1^{-1}\, \psi$.
\[  H_2 \,T_1^{-1}\psi = \left( h(S_L,S_1) + h\left(S_1, T_2 S_2  T_2^{-1}\right) + \sum_{j=3}^L h(S_{j-1}, S_j)\right) \,T_1^{-1} \psi = \]\be  T_1^{-1} \,\left( h(S_L,T_1 S_1 T_1^{-1}) + h\left(T_1 S_1 T_1^{-1}, T_2 S_2  T_2^{-1}\right) + \sum_{j=3}^L h(S_{j-1}, S_j)\right) \psi\,.  \ee
For $T_1^{-1}\, \psi$ to be an eigenvector  of $H_2$ the last expression should be equal to $E\, T_1^{-1}\psi $, but 
\be  E\, T_1^{-1}\psi =  T_1^{-1}\,H_1\psi =  T_1^{-1} \left(h(S_L,T_1 S_1 T_1^{-1}) + h\left( S_1 , \,S_2 \right) + \sum_{j=3}^L h(S_{j-1}, S_j)\right) \psi\ee
These are not generally equal unless
\be h\left(T_1 S_1  T_1^{-1}, T_2 S_2  T_2^{-1}\right) \;=\; h(S_1,S_2)\,,\ee  i.e.  $h(S_1,S_2)$ is $T$-invariant.
This leads to the conclusion that $T$ is only a true twist if $h(S_1,S_2)$  commutes with $T_1\,T_2$.

Now focusing on the XXZ model, for $T$ to be a true twist, we require that $T_1\,T_2$ commutes with the interaction term $h(S_1, S_2)$.   One can write this as
\be  h(S_1,S_2) =\frac{1}{2}\left(
S_1^{\rm x}\;S_{2}^{\rm x}+
S_1^{\rm y}\;S_{2}^{\rm y}+\frac{q\!+\!q^{-1}}{2}\;
(S_1^{\rm z}\;S_{2}^{\rm z}-1) \right)\ee
The eigenvalues of this operator are $\{-(1+q)^2/q,\; -(1-q)^2/q, \;0, \;0\}$; so the only freedom left for $T$ is transforming the eigenspace of the degenerate eigenvalue zero.   This leaves only
the complex generalizations of rotations around the z-axis.

One might object that the pure periodic Hamiltonian is unchanged, if one takes  $h(S_j,S_{j+1})= e_j$ given in (\ref{XXZmonoid}):  \be  h(S_1,S_2) \;=\; e_1 \;=\;\frac{1}{2}\left(S_1^{\rm x}\;S_2^{\rm x}+
S_1^{\rm y}\;S_2^{\rm y}+\frac{q\!+\!q^{-1}}{2}\;(
S_1^{\rm z}\;S_2^{\rm z}-1) +
\frac{q\!-\!q^{-1}}{2}(S_1^{\rm z}- S_2^{\rm z})\right)\ee
This has  eigenvalues $\{-q\!-\!q^{-1},0,0,0\}$.   Therefore it commutes with a greater family of operators, but still, demanding it to be of the form  $T_1 \,  T_2$ (a tensor product of two copies of $T$) leaves for $T$ only operators that commute with $S^{\rm z}$.

Clearly, a true twist for the XXZ model exists only in the form $T = \exp(f S^{\rm z})$, for some $f\in\C$.  

\BibSpec{article}{
+{}{\PrintAuthors} {author}
+{,}{ } {title}
+{,}{ \textit} {journal}
+{}{ \textbf} {volume}
+{}{ \parenthesize} {date}
+{,}{ } {pages}
+{,}{ } {note}
+{.}{} {transition}
+{}{ } {review}
}

\end{document}